\documentclass[aps,english,prl,floatfix,amsmath,superscriptaddress,tightenlines,twocolumn]{revtex4-2} %nofootinbib 

\usepackage[utf8]{inputenc}
\usepackage{graphicx}
\usepackage{amsmath, amssymb, amsfonts, amsthm, bm, nicefrac}
\usepackage{multirow} 
\usepackage{tabularx} 
\usepackage{array}
\usepackage{units}
\usepackage{tensor} 
\usepackage{braket}
\usepackage{hyperref}
\usepackage{dcolumn}
\usepackage{bm}
\usepackage{enumitem}
\usepackage{xcolor}
\usepackage{soul}
\usepackage[export]{adjustbox}
\usepackage{tikz}

\makeatletter 
\renewcommand\twocolumngrid{
	\def\footnoterule{
		\dimen@\skip\footins\divide\dimen@\thr@@
		\kern-\dimen@\hrule width.5in\kern\dimen@}
	\do@columngrid{mlt}{\tw@}
 }

\makeatother

\hypersetup{
    colorlinks=true,
    linkcolor=blue,
    filecolor=magenta,      
    urlcolor=blue,                                         %% 
    pdftitle={Geometric Additivity of Modular Commutator}, %% PDF의 제목
    pdfpagemode=FullScreen,
    }

\begin{document}

\title{Geometric Additivity of Modular Commutator for Multipartite Entanglement} 
\author{Sung-Min Park}
%\email{sungmin.park@kaist.ac.kr}
\affiliation{Department of Physics, Korea Advanced Institute of Science and Technology, Daejeon 34141, Korea}
\author{Isaac H. Kim \thanks{corresponding author}}
\email{ikekim@ucdavis.edu}
\affiliation{Department of Computer Science, University of California, Davis, CA 95616, USA}
\author{Eun-Gook Moon \thanks{corresponding author}}
\email{egmoon@kaist.ac.kr}
\affiliation{Department of Physics, Korea Advanced Institute of Science and Technology, Daejeon 34141, Korea}

\begin{abstract}
    A recent surge of research in many-body quantum entanglement has uncovered intriguing properties of quantum many-body systems. A prime example is the modular commutator, which can extract a topological invariant from a single wave function. Here, we unveil novel geometric properties of many-body entanglement via a modular commutator of two-dimensional gapped quantum many-body systems. We obtain the geometric additivity of a modular commutator, indicating that modular commutator for a multipartite system may be an integer multiple of the one for tripartite systems. Using our additivity formula, we also derive a curious identity for the modular commutators involving disconnected intervals in a certain class of conformal field theories. We further illustrate this geometric additivity for both bulk and edge subsystems using numerical calculations of the Haldane and $\pi$-flux models.
 \end{abstract}

\maketitle

\paragraph{Introduction:}
Two-dimensional gapped quantum systems exhibit intriguing many-body entanglement phenomena~\cite{sachdev2023quantum, wen2004quantum}. 
A remarkable aspect of these systems is the bulk-edge correspondence, which dictates that the effective field theory at the boundary determines the universal properties of the bulk.
A prime example is the quantum Hall effect in two spatial dimensions~\cite{klitzing1980new}, where a nonzero bulk Chern number indicates the presence of gapless edge modes~\cite{Hatsugai1993Chern, kitaev2006anyons}. 
These gapless edge modes can be characterized by the chiral central charge ($c_{-}$), which appears in the zero-temperature limit of the thermal Hall conductivity~\cite{kane1997quantized, read2000paired, kitaev2006anyons}. 
The bulk-edge correspondence suggests that the bulk ground state wave function may capture the chiral central charge, although how to extract it had remained puzzling.

\begin{figure}
    \centering
    \includegraphics[width=\columnwidth]{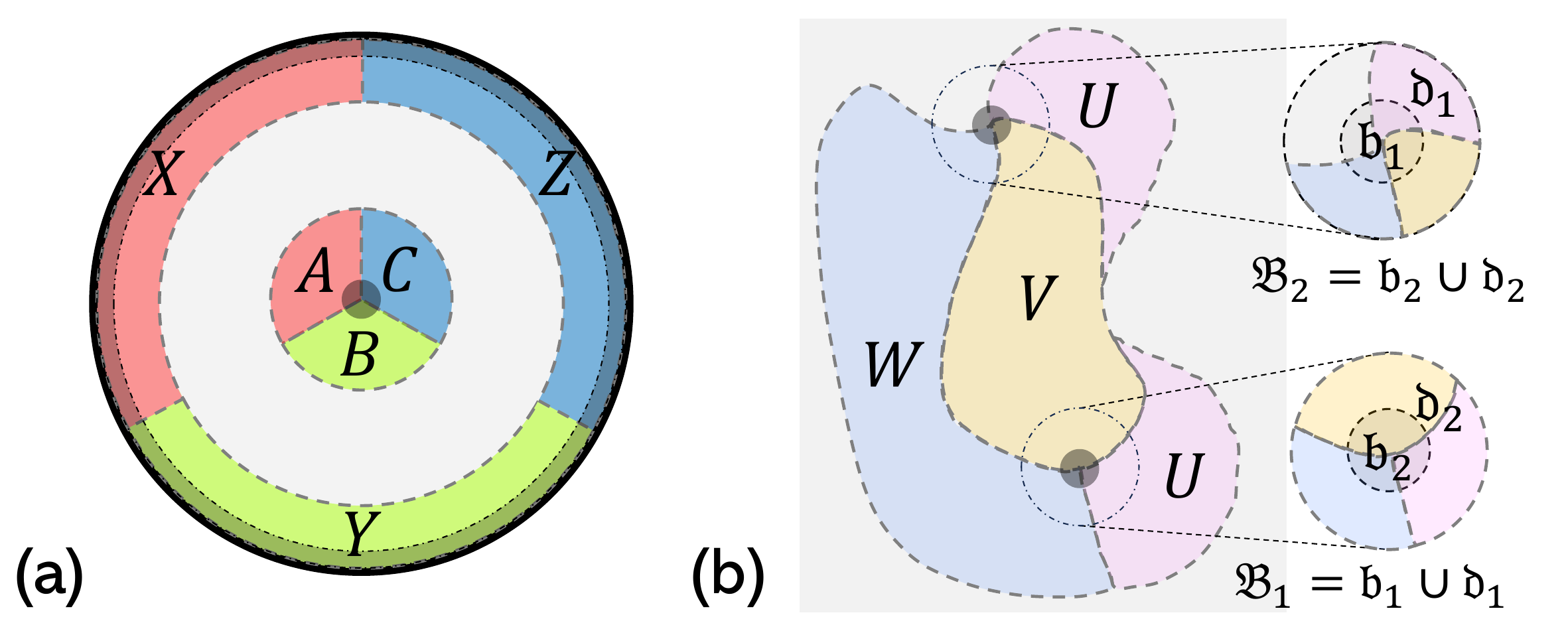}
    \caption{Exemplary partitions in two-dimensional systems.
    (a) Tripartite bulk disk $ABC$ and edge annulus $XYZ$.
    The shaded region represents the {\it physical} edge of the system.
    (b) Multipartite bulk partition with the subregions $U, V, W$.
    Two tri-junctions of ($U$, $V$, $W$) consist of two set of `balls', $\mathfrak{B}_{1} = \mathfrak{b}_{1} \cup \mathfrak{d}_{1}$ and $\mathfrak{B}_{2} = \mathfrak{b}_{2} \cup \mathfrak{d}_{2} $. 
    }
    \label{fig: Figure_1}
\end{figure}

Recently, Refs.~\cite{kim2022chiral, kim2022modular, zou2022modular, fan2022entanglement, fan2022generalized} demonstrated that the modular commutator of the ground state can capture the chiral central charge.
Let us recall the definition of a modular commutator
\begin{eqnarray}
    J(A,B,C)
    &\equiv& i \, \mathrm{Tr}( \rho_{ABC} [K_{AB}, K_{BC}] ), \nonumber 
\end{eqnarray}
where $K_{X} \equiv - \ln \rho_{X}$ is the modular Hamiltonian of the reduced density matrix $\rho_{X}$ on region $X$. 
For the ground state of a two-dimensional gapped system, $| \Psi \rangle$, the modular commutator gives the chiral central charge~\cite{kim2022chiral, kim2022modular, zou2022modular, fan2022entanglement, fan2022generalized}: 
\begin{equation}
    J(A, B, C)
    = \frac{\pi}{3} c_{-}.
    \label{eq: CCC}
\end{equation}
A tripartition $ABC$ with a complete tri-junction is illustrated in Fig.~\ref{fig: Figure_1}(a).
This result can be viewed as a part of an ongoing research program that aims to study universal properties of the underlying quantum many-body system via multipartite entanglement~\cite{Dehghani2021, zou2021universal, siva2022universal, fan2022extracting, Kobayashi2024extracting, kobayashi2024higher, liu2024anyon}. 
These recent developments call for further studies to explore the multipartite entanglement properties of many-body quantum systems.

In this paper, we extend the applicability of the modular commutator to more general geometries. 
Our work unveils a new universal geometric identity for the modular commutator. 
In particular, we find the curious identity that involves disconnected intervals in conformal field theories (CFTs) and singular regions in the bulk. 
The geometric aspects of the modular commutator of multipartite entanglement are illustrated by employing the area law of entanglement entropy~\cite{kitaev2006topological, levin2006detecting} and numerical calculations of lattice models.

\paragraph{Geometric Additivity:}
Originally, the modular commutator~\cite{kim2022chiral,kim2022modular} was defined over a region partitioned into three disks, all meeting at a single tri-junction. 
We consider a more general partition that goes beyond this original setup. 
An example of such a system is shown in Fig.~\ref{fig: Figure_1}(b), which features two tri-junctions. 
We find that the modular commutator for such subsystems can be expressed as:
\begin{eqnarray}
    J(U, V, W) = \sum_{i} J(U_{\mathfrak{B}_{i}}\!, V_{\mathfrak{B}_{i}}\!, W_{\mathfrak{B}_{i}}) + J(U_{\mathfrak{r}}, V_{\mathfrak{r}} , W_{\mathfrak{r}}),
    \label{eq: Geometric Additivity}
\end{eqnarray}
where an index $i$ specifies a tri-junction of a subregion $UVW$. 
At each tri-junction, $V_{\mathfrak{B}_{i}} = V \cap \mathfrak{B}_{i}$ is for the intersection between  a subregion $V$ and a ball $\mathfrak{B}_{i}$.  
For an each ball $\mathfrak{B}_{i}$, a smaller ball $\mathfrak{b}_{i}$ and an associated annulus region $\mathfrak{d}_{i}$ are introduced to zoom in a tri-junction, $\mathfrak{B}_{i} = \mathfrak{b}_{i} \cup \mathfrak{d}_{i}$. 
A relative complement region of all the tri-junctions is defined as $V_{\mathfrak{r}} = V \setminus (\cup_i \mathfrak{b}_{i})$, which contributes to the residual term, $J(U_{\mathfrak{r}}, V_{\mathfrak{r}}, W_{\mathfrak{r}})$. 
It then follows that $J(U_{\mathfrak{B}_{i}}, V_{\mathfrak{B}_{i}}, W_{\mathfrak{B}_{i}})$ is quantized as in Eqn.~\eqref{eq: CCC} for a complete junction, while an incomplete junction gives a non-quantized value. 
See Supplemental Materials (SM) for more details.

The modular commutator satisfies \emph{geometric additivity} if the residual term vanishes. 
We prove that the residual term indeed vanishes for invertible states by using properties of quantum Markov chains and the area law of the entanglement entropy~\cite{kitaev2006topological, levin2006detecting}. 
For non-invertible states (e.g., states that can host anyons), it is an open question whether the geometric additivity holds in general. 
We provide a conjecture from which the additivity would follow, though the proof of it remains open; see SM for more details.

\begin{figure}
    \centering
    \includegraphics[width=\columnwidth]{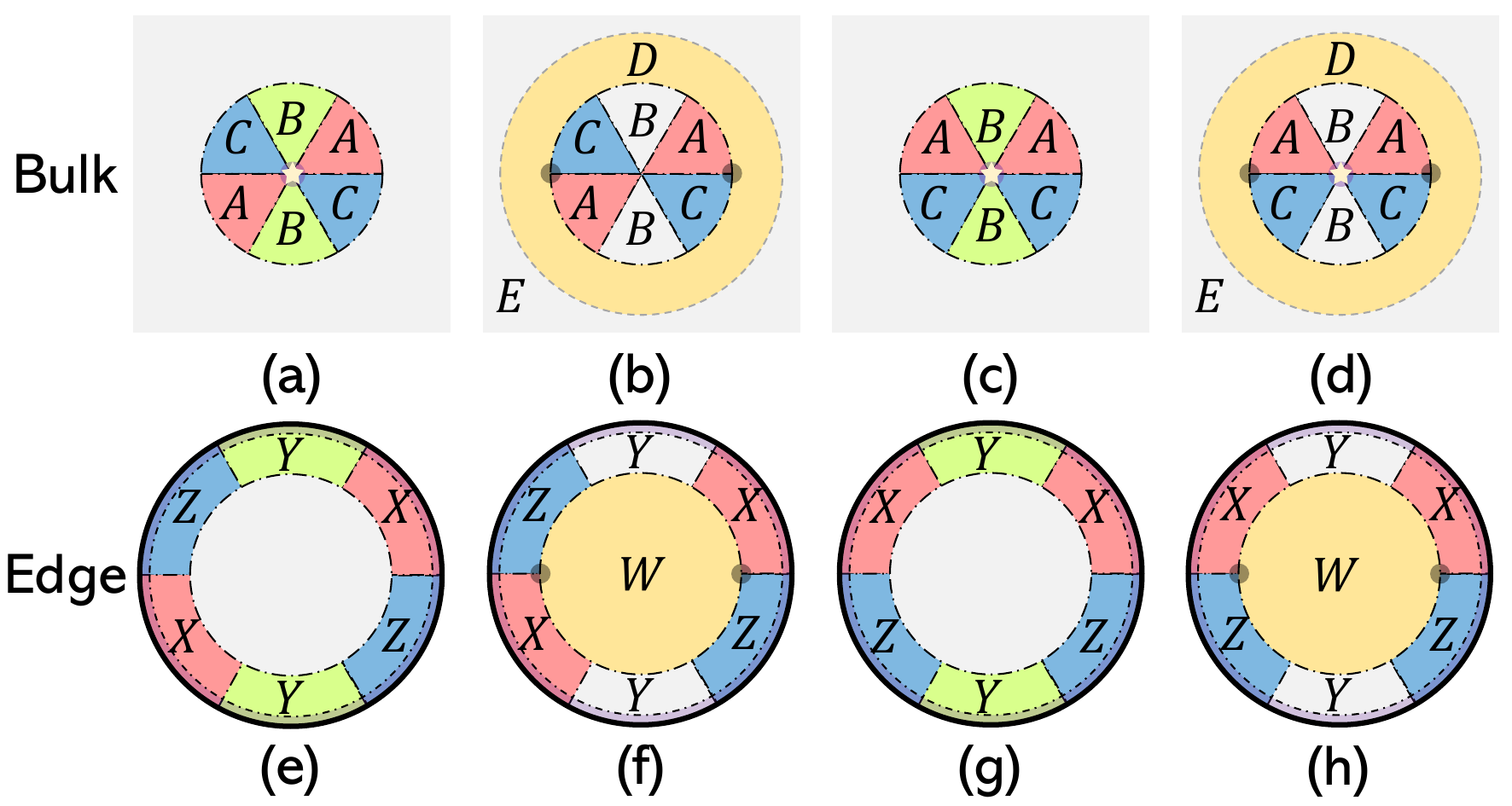}
    \caption{Bulk and edge pizza partitions and corresponding multipartite subsystems.
    (a),(c) The bulk pizza partitions of $ABC$. 
    (b),(d) The corresponding multipartite subsystems of (a) and (c), respectively.
    (e),(g) The edge pizza partitions of $XYZ$.
    (f),(h) The corresponding multipartite subsystems of (e) and (g), respectively.
    }
    \label{fig: Figure_2}
\end{figure}

\paragraph{Applications:}
One of the main applications of geometric additivity is an exact calculation of the modular commutator for new types of subsystems. As an example, we consider a partition in Fig.~\ref{fig: Figure_2}(a), which we call as the pizza partition.
Let us first consider a bulk system $CDA$ in Fig.~\ref{fig: Figure_2}(b) with the complementary property $K_{X}| \Psi \rangle = K_{\bar{X}} | \Psi \rangle$. 
The modular commutator for $CDA$ becomes
\begin{align}
    J(C,D,A) = &  i \langle \Psi | [ K_{ABE}, K_{BCE} ] | \Psi \rangle \nonumber \\
    = & i \langle \Psi | [ K_{AB}, K_{BC} ] | \Psi \rangle = J(A,B,C), 
    \label{eq: equivalence of J(C,D,A) and J(A,B,C)}
\end{align}
where in the second line, we use the fact that the state on the two sufficiently distant regions is a product state; for example, $K_{EAB} = K_E + K_{AB}$.
Thus, we can use $J(C, D, A)$ to determine $J(A, B, C)$. Using the additivity formula we obtain
\begin{equation}
    J(A,B,C) = 2 \times \frac{\pi}{3} c_{-}
\end{equation}
for Fig.~\ref{fig: Figure_1}(a). 
In essence, each complete tri-junction of $J(C, D, A)$ contributes to the modular commutator (by $\frac{\pi}{3} c_{-}$), yielding a result that is twice as large as the one for the tripartition in Fig.~\ref{fig: Figure_1}(a). 
Similarly, one can apply the additivity formula to the pizza partition in Fig.~\ref{fig: Figure_2}(c), and show $J(A, B, C) = 0$. 
This is because the contributions from the two tri-junctions cancel each other out.

The additivity formula may also be applied to the physical edge of the system. The similar use of complementary property gives $J(X, Y, Z) = J(Z, W, X)$, and one can apply the additivity to the complement region.
For example, for the pizza partition in Fig.~\ref{fig: Figure_2}(e), we find
\begin{eqnarray}
    J(X,Y,Z) = - 2 \times \frac{\pi}{3} c_{-}.
    \label{eq: edge pizza double}
\end{eqnarray}
We note that the modular Hamiltonians involved in the calculation of $J(X,Y,Z)$ are associated with disconnected intervals (e.g., $XY$ and $YZ$). 
This is very different from the setup considered in Ref.~\cite{zou2022modular}, which only involved intervals. 
While the entanglement Hamiltonian for an interval is local~\cite{cardy2016entanglement}, the entanglement Hamiltonian for disconnected intervals is not~\cite{arias2018entropy}. 
Our result shows that, in spite of this nonlocality, its modular commutator result in a substantially simpler form. 
However, our derivation only applies only to the edge theory of invertible states. 
It is currently unclear if Eqn.~\eqref{eq: edge pizza double} holds for any 1+1D CFTs.

% the edge region $XYZ$ forms disconnected intervals in CFTs~\cite{Tu2013, zou2022modular}, whose modular Hamiltonians are intrinsically non-local even in free theory~\cite{arias2018entropy}. 
% Our result suggests that, in spite of this nonlocality, its modular commutator may be substantially more simple. 
% It remains an open problem whether Eqn.~\eqref{eq: modular commutator in pizza partition} holds for any 1+1D CFTs. 

\begin{figure}
    \centering
    \includegraphics[width=\columnwidth]{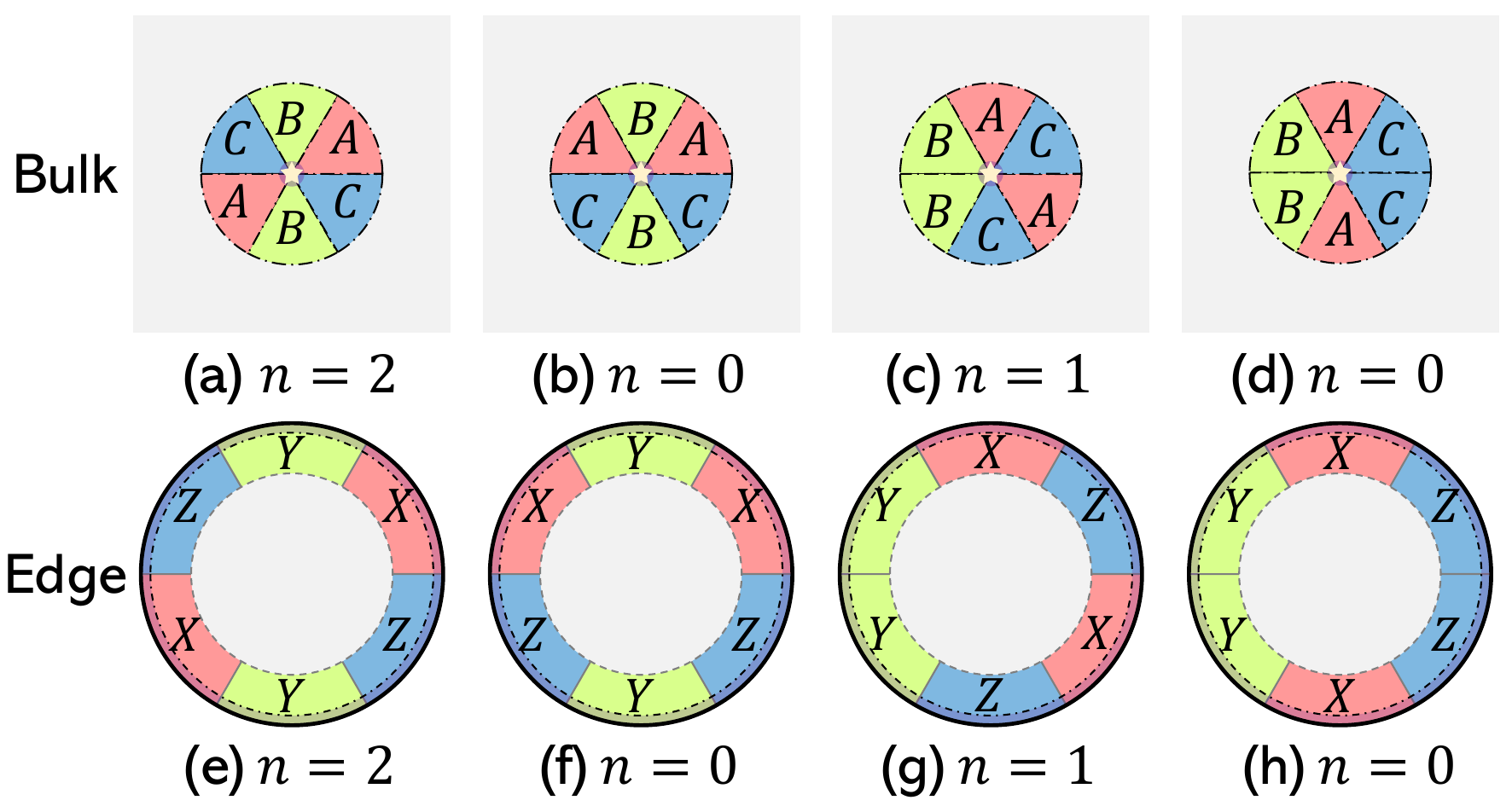}
    \caption{The geometric integer $n$ for the various bulk and edge pizza partitions.}
    \label{fig: Figure_3}
\end{figure}

The geometric additivity can be further applied to various pizza partitions as in Fig.~\ref{fig: Figure_3}.
It is straightforward to show that the modular commutators of a pizza partition yield
\begin{equation}
    J(A, B, C) = - J(X, Y, Z) = \frac{\pi}{3}  \big( c_{-} \times n  \big), \quad n \in \mathbb{Z},
    \label{eq: modular commutator in pizza partition}
\end{equation}
where the arrangement of the bulk and edge pizza partition, $ABC$ and $XYZ$, are topologically equivalent.
The chiral central charge $c_{-}$ is determined by a ground state $| \Psi \rangle$, and the integer $n$ depends on the arrangement near the tri-junction. 
We refer to it as the \textit{geometric integer} in this work.
We conjecture that Eqn.~\eqref{eq: modular commutator in pizza partition} holds for most two-dimensional gapped systems, although our proof is limited to invertible states.

Lastly, we discuss the application of the additivity formula for incomplete junctions. 
An incomplete tri-junction generally exhibits non-quantized values dependent on microscopic details~\cite{fan2022entanglement}. 
Yet, we find that the modular commutator with an incomplete junction has intriguing complementary properties.
Namely, for a disk-like region $ABCD$, the sum of two modular commutators with incomplete junctions yields the following complementary relation:
\begin{equation}
    \includegraphics[width=1.25cm, valign=c]{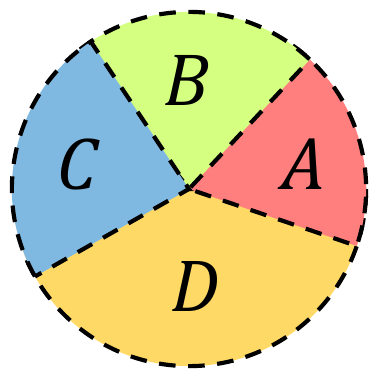} \;\;\;\;\;\;
    J(A,B,C) + J(B, C, D) = \frac{\pi}{3} c_{-}, 
\end{equation}
where the subregion $A, B, C, D$ meet at a point. 
This is an immediate consequence of geometric additivity.
Interestingly, this identity holds even for non-invertible states, i.e., topologically ordered states that can host anyons; see the SM. Note that this complementary relation can be viewed as a bulk analog of the CFT identity recently discovered in Refs.~\cite{zou2022modular, kim2024conformal, fan2022entanglement}.
A particular application of this identity arises when the system exhibits proper spatial symmetry near its tri-junction. 
This allows us to extract the chiral central charge within a smaller system size, yielding a \textit{half}-quantized value:  $J(A, B, C)_{\sigma} = J(B, C, D)_{\sigma} = \frac{1}{2} \times \frac{\pi}{3} c_{-}$.

\begin{figure}[!t]
    \begin{center}
        \includegraphics[width=\columnwidth]{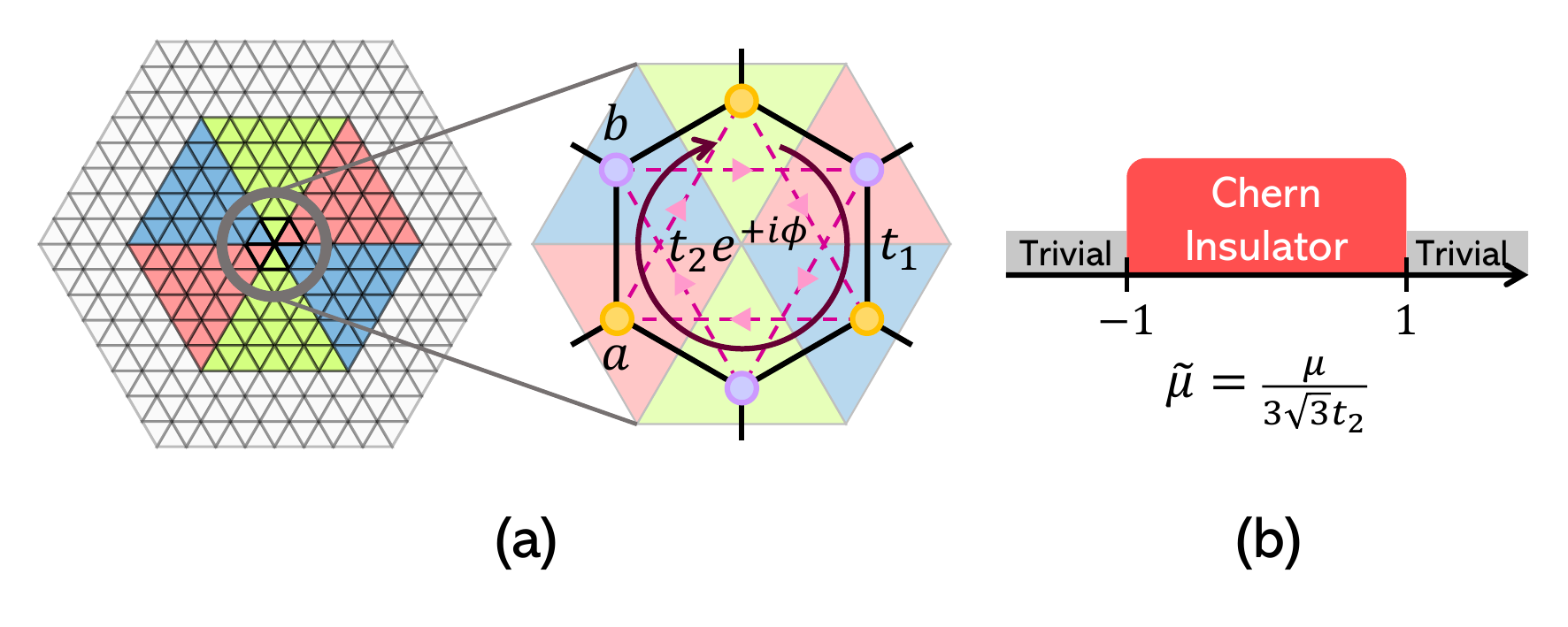}
        \caption{Schematic description of the Haldane model
        (a) Upside-down triangles ($\bigtriangledown$) and triangles ($\bigtriangleup$) represent sublattices $a$ and $b$, respectively.
        See the main text for the information on hopping parameters. 
        (b) The phase diagram of the Haldane model.
        In the chern insulator phase at $-1< \tilde{\mu} <1$, chiral central charge is $c_{-}=1$, otherwise zero.
        }
        \label{fig: Figure_4}
    \end{center}
\end{figure}
 
\paragraph{Numerical calculation:}
We numerically verify the geometric additivity for certain lattice models. We consider the Haldane model on the honeycomb lattice~\cite{haldane1988model}, whose Hamiltonian consists of three parts:
\begin{equation}
    H_{H} = H_{0} (\mu) + H_{1} (t_{1}) + H_{2} (t_{2}; \phi). \nonumber
\end{equation}
The first term contains the on-site energy terms with strength $\mu$. 
We set the onsite energy to be $\mu$ for $a$ sublattice and $-\mu$ for $b$ sublattice: $H_{0}(\mu) = \mu \sum_{r}  (c^{\dagger}_{r,a} c_{r,a} - c^{\dagger}_{r,b} c_{r,b})$. 
The second term $H_{1}$ includes the nearest-neighbor hopping terms with amplitude $t_{1}$, which are represented as black solid links in Fig.~\ref{fig: Figure_4}(a): $H_{1}(t_{1}) = t_{1} \sum_{\langle j, k \rangle} c^{\dagger}_{j} c_{k}$. 
The last term has the next nearest neighbor hopping terms with complex amplitude $t_{2}$ with phase $\phi$, which breaks the time-reversal symmetry: $H_{2}(t_{2};\phi) = |t_{2}| \sum_{\langle \langle j,k \rangle \rangle} e^{-i \phi \nu_{jk}} c^{\dagger}_{j} c_{k}$. Here, $\nu_{jk}$ are $\pm 1$ and depends on the arrow's direction. 
The dashed arrows in Fig.~\ref{fig: Figure_4}(a) denote the next nearest hopping. The hopping in the direction of the arrow accumulates the flux $\phi$.
Setting a tuning parameter $\tilde{\mu}=\mu/(3 \sqrt{3} t_{2})$ with $ t_{1} = t_{2} = 1$, we choose a one-dimensional path along $\phi = \pi/2$ where two topological phase transitions are present as in Fig.~\ref{fig: Figure_4}(b).  It is well known that the Haldane model has the chiral central charge $c_{-} = 1$ for $-1 < \tilde{\mu}<1$ and $c_{-} =0$ for $\tilde{\mu}>1$ and $\tilde{\mu}<-1$.

To evaluate its modular commutator, we divide the bulk lattice system into six sectors for a pizza partition. 
For example, a lattice realization of the bulk $ABC$ is shown in Fig.~\ref{fig: Figure_4}(a).
Note that the total number of lattice points in the pizza partition is $867$, which makes each sector large enough except at the critical points at $\tilde{\mu}=\pm 1$.

\begin{figure}[!t]
    \begin{center}
        \includegraphics[width=\columnwidth]{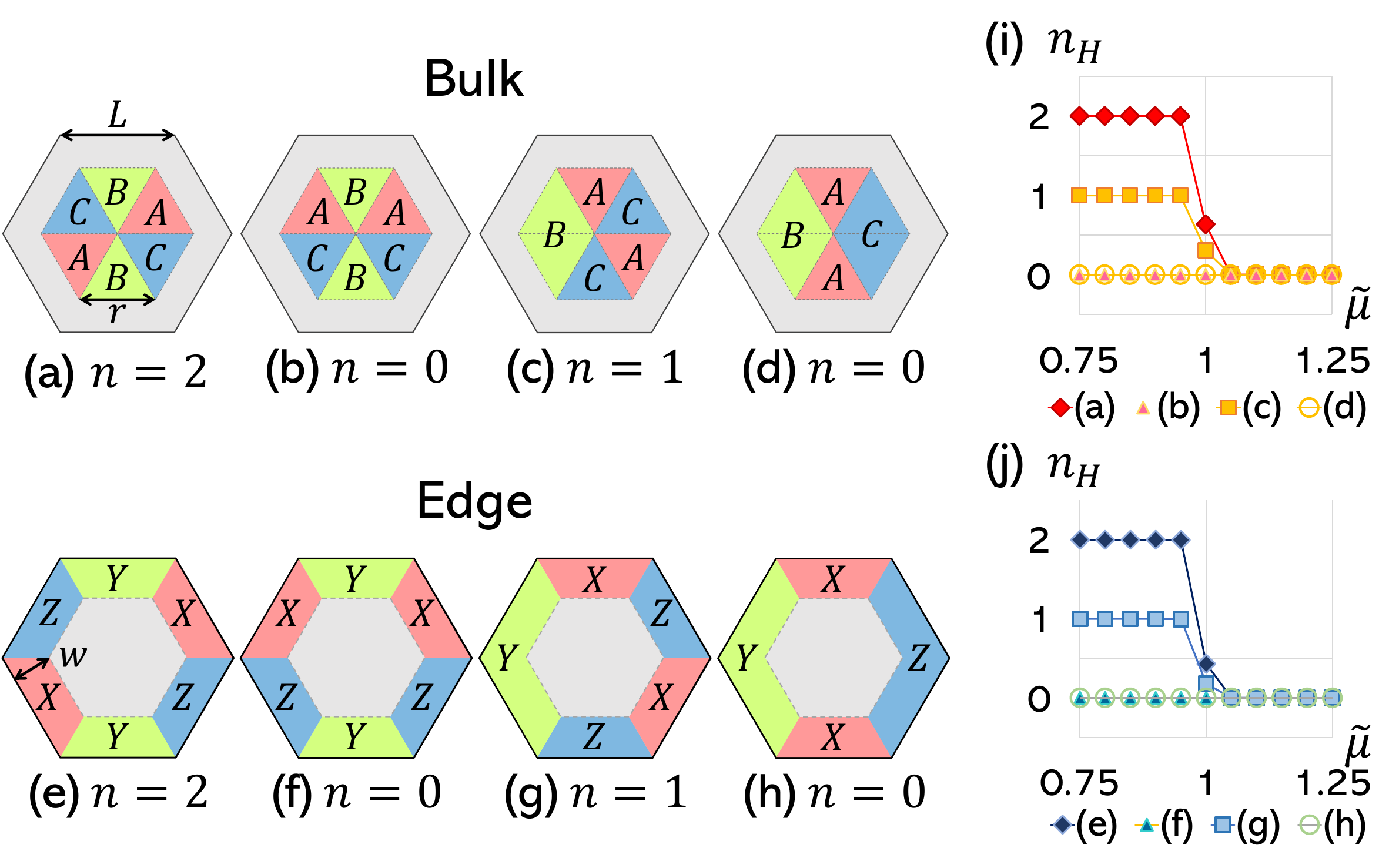}
        \caption{Numerical evaluation of the geometric integer ($n_{H}$) for the pizza partitions.
        (a)-(h) The lattice realizations of the pizza partitions. We fix the linear system size $L$, bulk subsystem size $r$, and the edge width $w$: $L = 24$,  $r=17$, and $w = 13$. 
        (i) Plot of $n_{H}$ for the bulk pizza partitions with the tuning parameter $\tilde{\mu}$.
        (j) Plot of $n_{H}$ for the edge pizza partitions with the tuning parameter $\tilde{\mu}$.}
        \label{fig: Figure_5}
    \end{center}
\end{figure}

The exact ground state of the Haldane model, $| \psi_H \rangle$, is readily obtained with a given lattice, and its modular commutator is numerically evaluated,
\begin{eqnarray}
    n_{H} \equiv \frac{3}{\pi}  J(A,B,C)= i \frac{3}{\pi} \langle \psi_H | [K_{AB}, K_{BC}] | \psi_H \rangle. 
\end{eqnarray}
The numerical values of $n_{H}$ for the bulk pizza partitions, Fig.~\ref{fig: Figure_5}(a)-(d), are illustrated in Fig.~\ref{fig: Figure_5}(i) while the ones of  $n_{H} =-\frac{3}{\pi}  i \langle \psi_H | [K_{XY}, K_{YZ}] | \psi_H \rangle $ for the edge pizza partitions, Fig.~\ref{fig: Figure_5}(e)-(f), are illustrated in Fig.~\ref{fig: Figure_5}(j).
We emphasize that the numerical calculations of $n_{H}$ are precisely matched with the results of the geometric additivity away from the critical points ($\tilde{\mu}=\pm1$).

To check the validity of our results, we also introduce small disorder to our numerical calculations by adding the  Anderson term~\cite{anderson1958absence}, $\sum_{j} V_{j} c^{\dagger}_{j} c_{j}$, where  $V_{j}$ is sampled from a uniform distribution within the range $[-W/2, W/2]$. 
We choose $W$ as 10\% of the bulk energy gap, and no differences in the numerical calculations are found. 
We also perform similar calculations for the $\pi$-flux model on the square lattice and obtain qualitatively similar results for the higher geometric integers $n_{\pi}=3, 4$. 
These results are presented in the SM.

\paragraph{Discussion and conclusion:}

Intuitively, the geometric integer $n$ can be understood in terms of the modular current~\cite{kitaev2006anyons, kim2022chiral, kim2022modular}. 
For any two regions $L$ and $R$, this is defined as $f(L, R) = \sum_{v \in L} \sum_{u \in R} \tilde{f}_{vu}$, 
where $\tilde{f}_{vu} = i \langle [ \tilde{K}_{v} , \tilde{K}_{u}]  \rangle$ and $\tilde{K}_{v}$ is the local modular Hamiltonian such that $K_{X} = \sum_{v \in X} \tilde{K}_{v}$. 
It is well understood that non-trivial modular current flows along the boundaries of nearby regions~\cite{kim2022chiral,kim2022modular}.

\begin{figure}[]
    \centering
    \includegraphics[width=\columnwidth]{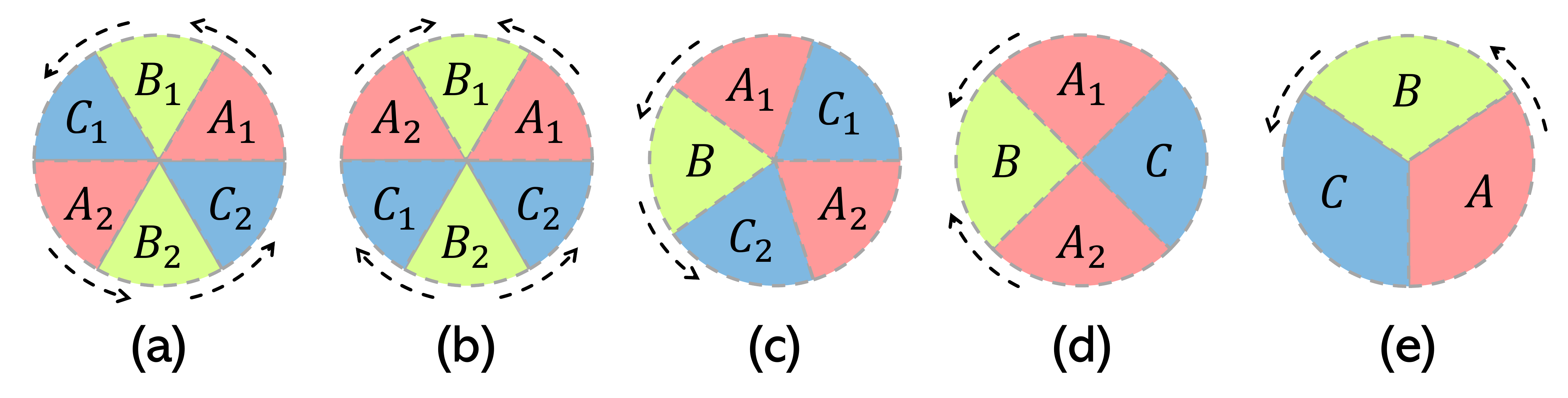}
    \caption{The modular current interpretation for the pizza partitions. 
    The non-trivial modular current of $J(A, B, C)$ flows from the subregions of $AB$ into the subregions $BC$ in the vicinity of their boundaries.
    (a)-(d) The pizza partitions of Fig.~\ref{fig: Figure_2}. 
    (e) The pizza partition of the typical tri-junction as in Fig.~\ref{fig: Figure_1}(a).
    }
    \label{fig: Figure_6}
\end{figure}

We can apply this intuition to the bulk pizza partitions, reproducing the results we have shown rigorously. 
Due to the locality of the bulk modular Hamiltonian, the modular commutator may be rewritten as
\begin{equation}
    J(A, B, C) =  \sum_{v \in AB} \sum_{ u \in BC} \tilde{f}_{vu}. \nonumber
\end{equation}
Thus, the modular current can be evaluated as a sum of current flowing from a subregion of $AB$ to another subregion of $BC$. 
For instance, consider the partition of Fig.~\ref{fig: Figure_6}(a), where the subregion is written as  $ABC = A_{1}B_{1}C_{1} A_{2}B_{2}C_{2}$. The modular current has contributions from the four boundaries,  $(A_{1}, B_{1})$, $(B_{1}, C_{1})$,  $(A_{2}, B_{2})$, $(B_{2}, C_{2})$. 
Thus, the geometric integer is twice as large as the one of Fig.~\ref{fig: Figure_6}(e). 
We can apply a similar reasoning to the other pizza partitions, which yields results consistent with our analysis based on the geometric additivity.

To conclude, we posit a new geometric property of many-body quantum entanglement in two-dimensional gapped quantum many-body systems. 
This is the geometric additivity of the modular commutator, which indicates that modular commutator for subsystems more general than the one considered in Ref.~\cite{kim2022chiral,kim2022modular,zou2021universal} can be decomposed into a sum of modular commutators over simplified subsystems (e.g., involving balls or intervals). 
Numerical calculations of the Haldane model corroborate the geometric additivity, both in the bulk and at the edge.

We remark that a recent work~\cite{gass2024many} reported spurious contributions of modular commutators.
In our numerical calculations, we have not observed such contributions, which is consistent with the previous work on the chiral central charge of free fermion models ~\cite{fan2022generalized}. 
In future works, it would be desirable to develop a different method for extracting the chiral central charge which is free of such problems. 
One interesting question is whether the additivity can be proved rigorously for free fermion models.

Another important question is whether the geometric additivity holds for non-invertible states. 
We note that the additivity can be proved for a certain (non-vacuum) reduced density matrix that contains anyons; see the SM. 
However, how to relate this result to the state of interest, i.e., the vacuum-reduced density matrix, is unclear at the moment.

% We note that the geometric additivity holds even for non-invertible states provided that the modular commutator is invariant in the presence of anyons; see the SM. 
%While it is tempting to accept the assumption, its validity needs to be rigorously checked in future works. \IK{I will revisit this part after checking the SM.}

\paragraph{Acknowledgement:}
We thank K. Hwang, J. McGreevy, X. Li, T.-C. Lin, and B. Shi for helpful discussions. IK acknowledges support from NSF under award number PHY-2337931. 
S.-M.P. and E.-G.M. were supported by 2021R1A2C4001847, 2022M3H4A1A04074153, National Measurement Standard Services and Technical Services for SME funded by Korea Research Institute of Standards and Science (KRISS – 2024 – GP2024-0015) and the Nano \& Material Technology Development Program through the National Research Foundation of Korea(NRF) funded by Ministry of Science and ICT(RS-2023-00281839).

\bibliography{citation}

%%%%%%%%%%%%%%%%%%%%%%%%%%%%%%%%%%%%%%%%%%%%%%%%%%%%%%%%%%%%%%%%%%%%%%%%%%%%
%%%%%%%%%%%%%%%%%%%%%%%%%%%%%%%%%%%%%%%%%%%%%%%%%%%%%%%%%%%%%%%%%%%%%%%%%%%%
%%%%%%%%%%%%%%%%%%%%%    SUPPLEMENTAL MATERIALS     %%%%%%%%%%%%%%%%%%%%%%%%
%%%%%%%%%%%%%%%%%%%%%%%%%%%%%%%%%%%%%%%%%%%%%%%%%%%%%%%%%%%%%%%%%%%%%%%%%%%%
%%%%%%%%%%%%%%%%%%%%%%%%%%%%%%%%%%%%%%%%%%%%%%%%%%%%%%%%%%%%%%%%%%%%%%%%%%%%

\newpage
\appendix

\tableofcontents

\vspace{1cm}

The Supplementary Material is organized as follows. 
Section \ref{sec: Geometric additivity of a modular commutator} provides the basic setup and a derivation of the geometric additivity formula. 
We provide another type of additivity formula for the incomplete disk in Section \ref{sec: Incomplete disk}. 
Section \ref{Sec: Numerical Simulation on pi-flux} presents results of numerical calculations on the $\pi$-flux model, including incomplete disk and $n \ge 3$ disconnected edge intervals. In Section \ref{sec: absence of the residual term}, we discuss the modular commutator of the pizza partition for invertible states and also provide remarks for the non-invertible states.

\section{Geometric additivity of a modular commutator}\label{sec: Geometric additivity of a modular commutator}

In this Section, we derive the geometric additivity formula using the area law of entanglement entropy~\cite{kitaev2006topological, levin2006detecting}, or equivalently, the entanglement bootstrap axiom \textbf{A1}~\cite{shi2020fusion}.

\subsection{Setup: area law of the entanglement entropy}\label{subsec: Entanglement Bootstrap Axioms}

\begin{figure}[!b]
    \centering
    \includegraphics[width = 0.9\columnwidth]{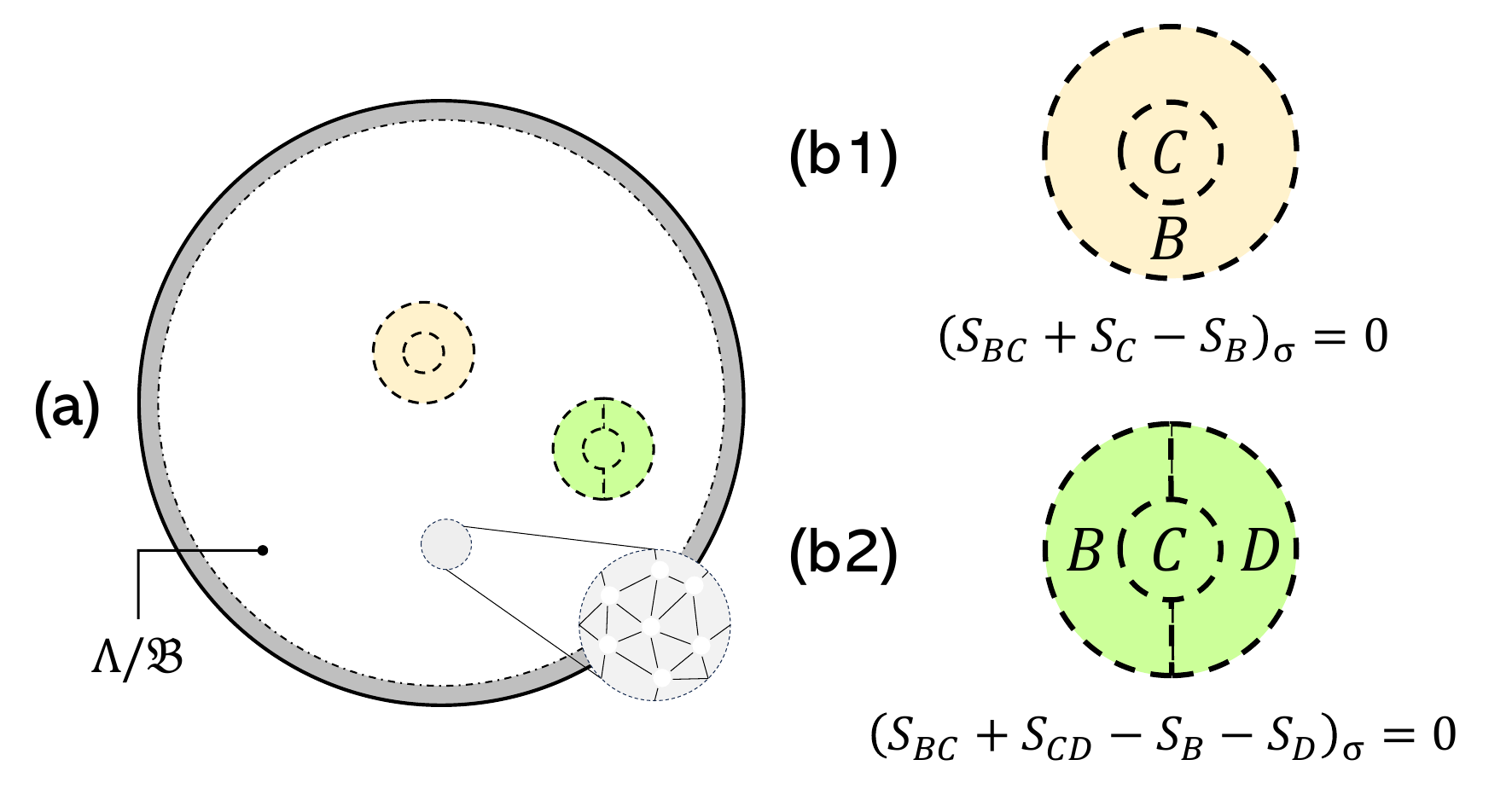}
    \caption{Schematic descriptions on the two entropic axioms.
    (a) A gapped quantum many-body system $\Lambda$, and the bulk `ball' regions $\mathfrak{B}$ lying in the bulk. 
    The solid lines at the edge represent the physical boundary, and the dashed lines represent the entanglement cut.
    (b) Two entropic combinations under area law.
    Subregions $B$, $C$, and $D$ are sufficiently larger than the bulk correlation length.} 
    \label{fig: SM_1}
\end{figure}

The entanglement bootstrap~\cite{shi2020fusion} starts with two axioms on the entanglement entropy of the bulk ball $\mathfrak{B}$ [Fig.~\ref{fig: SM_1}(b)]:
\begin{subequations}
    \begin{eqnarray}
        \mathbf{A0}\! : \; & \; 
        \Delta(B,C)_{\sigma}     = 0,  \label{eq: Axiom A0} \\
        \mathbf{A1}\! : \; & \; 
        \Delta(B, C, D)_{\sigma} = 0,  \label{eq: Axiom A1} 
    \end{eqnarray}
    \label{eq: The entanglement bootstrap axioms}
\end{subequations}
where the subscripts $\sigma$ of the parentheses is a state satisfying the axioms~\footnote{Intuitively, it might be convenient to view this state as a ground state of a gapped Hamiltonian, though we do not make use of that fact.}
and $\Delta$s are linear combinations of entanglement entropies:
\begin{subequations}
    \begin{eqnarray}
        \Delta(B,C)_{\rho} & := \! & ( S_{BC} + S_{C} - S_{B} )_{\rho} \ge 0 ,           \label{eq: Araki-Lieb} \\
        \Delta(B, C, D)_{\rho} & := \! & ( S_{BC} + S_{CD} \! - S_{B} \! - S_{D} )_{\rho} \! \ge \! 0 .  \label{eq: weak monotonicity} 
    \end{eqnarray}
    \label{eq: two entropy combinations}
\end{subequations}
Here, the subscript represents the relevant subsystem. For instance, $(S_{BC} + \ldots)_{\rho}$ is a shorthand notation for $S(\rho_{BC}) +\ldots$, where $S(\rho_{BC}) = -\text{Tr}(\rho_{BC} \ln \rho_{BC})$ is the von Neumann entropy. 
Note that two entropic combinations are always non-negative by the Araki-Lieb inequality and strong subadditivity (SSA).

An immediate consequence of Eqns.~\eqref{eq: The entanglement bootstrap axioms} is that for \textit{any} subsystem $A \subset \Lambda\setminus\mathfrak{B}$, $\sigma_{ABC}$ forms a quantum Markov chain (with respect to the choice of $B$ and $C$ appearing in Eqns.~\eqref{eq: two entropy combinations}). In other words, the conditional mutual information of a tripartite state $\sigma_{ABC}$ --- defined as $I(A: C|B)_{\rho}:= (S_{AB}+S_{BC} - S_{B} - S_{ABC})_{\rho}.$ --- vanishes:
\begin{eqnarray}
    I(A: C|B)_{\sigma} = 0.
\end{eqnarray}
Moreover, the modular Hamiltonian ($K_{X} \equiv - \ln \rho_{X}$) for a quantum Markov states $\sigma_{ABC}$ can be decomposed into the following form~\cite{petz2003monotonicity}:
\begin{equation}
    I(A:C|B) = 0\Longleftrightarrow
    K_{ABC} = K_{AB} + K_{BC} - K_{B}.
    \label{eq: the Petz's relation}
\end{equation}
By repeatedly applying the same argument, one can often decompose the modular Hamiltonian as a linear combination of terms that act on increasingly smaller subsystems~\cite{kim2022modular}.

\subsection{Derivation: geometric additivity formula}

In this subsection, we prove the geometric additivity of the modular commutator,
\begin{equation}
    J(U, V, W) \!
    = \! \sum_{i} J(U_{\mathfrak{B}_{i}}\!,  V_{\mathfrak{B}_{i}}\!, W_{\mathfrak{B}_{i}}) 
    + J(U_{\mathfrak{r}}, V_{\mathfrak{r}},  W_{\mathfrak{r}}),
    \label{eq: Sum over tri-junctions} 
\end{equation}
using entanglement bootstrap axiom \textbf{A1} [Eqn.~\eqref{eq: Axiom A1}].
Throughout this subsection, we will use the notation summarized in Table.~\ref{tab: notation summary} and suppress the notation for $\sigma$ unless it is unclear.

\begin{figure}[!b]
    \begin{center}
        \includegraphics[width=\columnwidth]{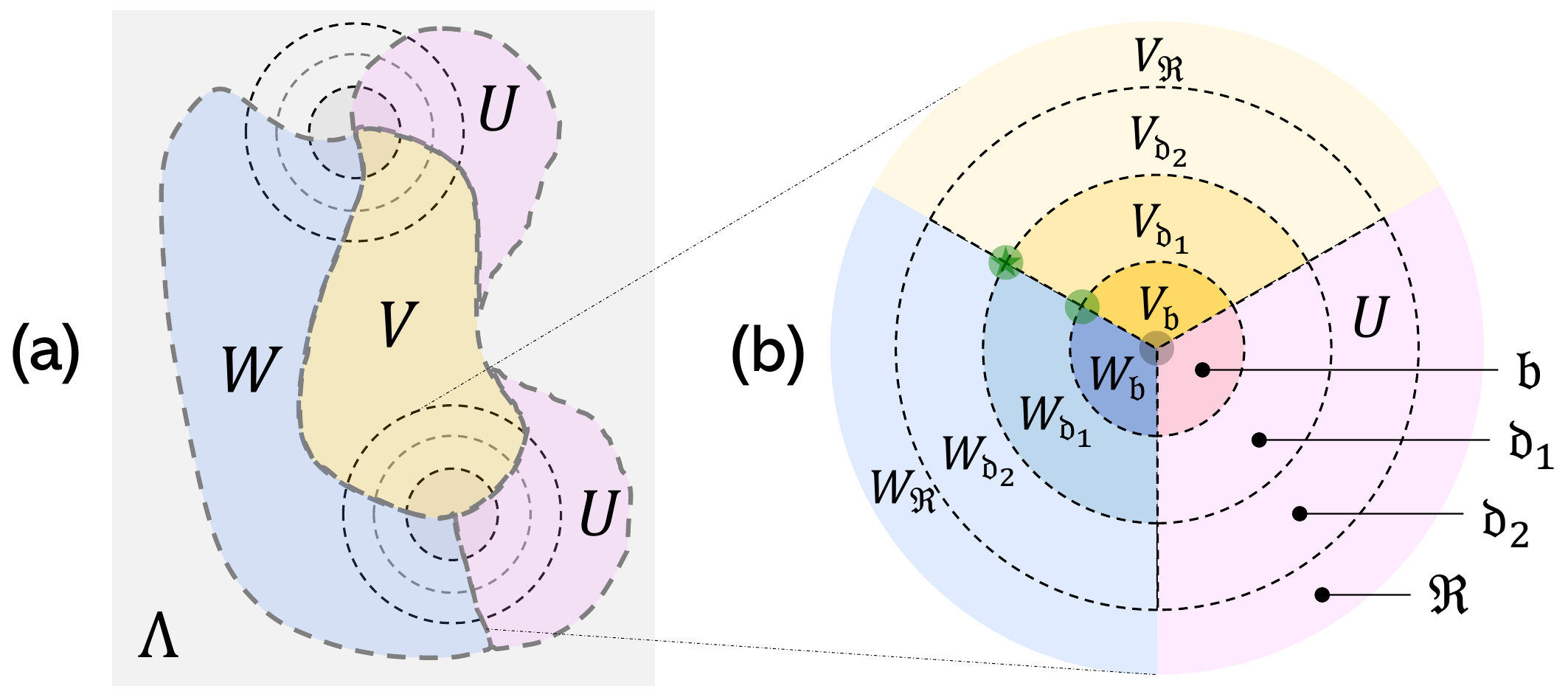}
        \caption{
        (a) Examplary multi-partition $UVW$ in $UVW$ in two-dimensional quantum many-body system $\Lambda$.
        (b) Schematic description of the $i$'th ball region, $\mathfrak{B}_{i} = \mathfrak{b}_{i} \cup \mathfrak{d}_{i}$. 
        The black and green shaded region are tri-junctions in Eqn.~\eqref{eq: Modular commutator J(U, V, W) - 1}.}
        \label{fig: SM_2}
    \end{center}  
\end{figure}

For concreteness,  we consider an exemplary multipartite region $U$, $V$, and $W$ [Fig.~\ref{fig: SM_2}(a)]. This will be our guiding example for proving the additivity formula. We note that these specific regions are chosen only for pedagogical purposes. The proof itself can be generalized straightforwardly to other choices of subsystems. This is because our approach is to sequentially excise a ball that includes a tri-junction one by one; for any choice of subsystems, one can simply repeat the same procedure, arriving at the additivity formula [Eqn.~\eqref{eq: Sum over tri-junctions}].

\begin{table}[]
    \centering
    \begin{tabular}{c|c}
        \hline
        \hline
        $\mathfrak{b}_{j}$ & a ball centered at $j$-th tri-junction  \\
        $\mathfrak{d}_{j}$ & an annulus surrounding $\mathfrak{b}_{j}$, $\mathfrak{d}_{j} = \mathfrak{d}_{1,j} \cup \mathfrak{d}_{2,j}$ \\
        $\mathfrak{B}_{j}$ & Union of $\mathfrak{b}_{j}$ and $\mathfrak{d}_{j}$, $\mathfrak{B}_{j} = \mathfrak{b}_{j} \cup \mathfrak{d}_{j}$ \\
        $\mathfrak{r}_{j}$ & The relative complement of $\mathfrak{b}_{j}$, \textit{e.g.} $\mathfrak{r}_{j} = \Lambda/\mathfrak{b}_{j}$ \\
        $\mathfrak{R}_{j}$ & The relative complement of $\mathfrak{B}_{j}$, \textit{e.g.} $\mathfrak{R}_{j} = \Lambda/\mathfrak{B}_{j}$ \\
        \hline
        \hline
    \end{tabular}
    \caption{Summary of notations (See also Fig.~\ref{fig: SM_2}(b))}
    \label{tab: notation summary}
\end{table}

The following are the key observations.
First, we can apply axiom \textbf{A1} to a ball containing the tri-junction and thus decompose the modular Hamiltonian of $UV$ and $VW$ near the tri-junction [Eqn.~\eqref{eq: the Petz's relation}].
For instance, consider the $i$-th tri-junction of the subregion $VW$ and apply the axiom \textbf{A1}.
Without loss of generality, we can decompose the modular Hamiltonian of $VW$ near the tri-junction, thanks to Eqn.~\eqref{eq: def of bd region}:
\begin{equation}
    I((VW)_{\mathfrak{b}_{i}}: (VW)_{\mathfrak{d}_{2, i}\mathfrak{R}_{i}} | (VW)_{\mathfrak{d}_{1, i}}) = 0,
    \label{eq: def of bd region}
\end{equation}
where $\mathfrak{d}_{1, i}$ is the inner annulus surrounding $\mathfrak{b}_{i}$, and $\mathfrak{d}_{2, i}$ is the outer annulus surrounding $\mathfrak{d}_{1, i}$ [Fig.~\ref{fig: SM_2}(b)].
The subscripts of $VW$ denote the intersections.
The second observation is that applying this decomposition yields contributions from additional tri-junctions, which exactly cancel each other out; see the two green-shaded regions in Fig.~\ref{fig: SM_2}(b).

We now elaborate on these observations in more detail, focusing on a tri-junction of $UVW$.
In what follows, we will obtain a succinct expression [Eqn.~\eqref{eq: Modular commutator J(U, V, W) - 4}] for $J(U, V, W)$. 
We start by rewriting the modular commutator as
\begin{eqnarray}
    J(U, V, W) 
    & = i \langle [ K_{UV}, K_{(VW)_{\mathfrak{bd}_{1}}} ] \rangle & - i \langle [ K_{UV}, K_{(VW)_{\mathfrak{d}_{1}}}] \rangle \nonumber  \\ 
    & + i \langle [ K_{UV}, K_{(VW)_{\mathfrak{r}}}] \rangle.   & 
    \label{eq: Modular commutator J(U, V, W) - 1}
\end{eqnarray}
All three terms are generally nonzero since they all have tri-junctions [green shaded regions in Fig.~\ref{fig: SM_2}(b)]. 
However, we can judiciously cancel out some of those contributions using the quantum Markov chain, such as Eqn.~\eqref{eq: def of bd region}.

Let us compute the first line of Eqn.~\eqref{eq: Modular commutator J(U, V, W) - 1}.
\begin{eqnarray}
      & \langle [ K_{UV}, K_{(VW)_{\mathfrak{bd}_{1}}}] \rangle - &  \langle [ K_{UV}, K_{(VW)_{\mathfrak{d}_{1}}}] \rangle   \nonumber \\
    = & \langle [ K_{(UV)_{\mathfrak{B}}}, K_{(VW)_{\mathfrak{bd}_{1}}} & - K_{(VW)_{\mathfrak{d}_{1}}} ] \rangle \nonumber \\
    = & \langle [ K_{(UV)_{\mathfrak{B}}}, K_{(VW)_{\mathfrak{B}}} ] \rangle & - \langle [ K_{(UV)_{\mathfrak{B}}}, K_{(VW)_{\mathfrak{d}}}] \rangle,
    \label{eq: Modular commutator J(U, V, W) - 2}
\end{eqnarray}
where in the second line of Eqn.~\eqref{eq: Modular commutator J(U, V, W) - 2} we reduce the region $UV$ into its restriction onto the ball $\mathfrak{B}$, by using an identity $\langle [ K_{UV}, K_{(VW)_{\mathfrak{bd}_{1}}}] \rangle = \langle [ K_{(UV)_{\mathfrak{B}}}, K_{(VW)_{\mathfrak{bd}_{1}}}] \rangle$. Using a similar identity, we extend the region $\mathfrak{d}_{1}$ of $VW$ to $\mathfrak{d} = \mathfrak{d}_{1} \cup \mathfrak{d}_{2}$ in the third line.

The second line of Eqn.~\eqref{eq: Modular commutator J(U, V, W) - 1} can be also calculated similarly:
\begin{eqnarray}
    \langle [ K_{UV}, K_{(VW)_{ \mathfrak{r} }} ] \rangle 
    = & \langle [ K_{(UV)_{\mathfrak{bd}_{1}}} - K_{(UV)_{\mathfrak{d}_{1}}}, K_{(VW)_{\mathfrak{r}}} ] \rangle \nonumber  \\ 
    & + \langle [ K_{(UV)_{\mathfrak{r}}}, K_{(VW)_{\mathfrak{r}}} ] \rangle \nonumber \\
    = & \langle [ K_{(UV)_{\mathfrak{B}}} - K_{(UV)_{\mathfrak{d}}} , K_{(VW)_{\mathfrak{d}}} ] \rangle \nonumber \\
    & + \langle [ K_{(UV)_{\mathfrak{r}}}, K_{(VW)_{\mathfrak{r}}} ] \rangle,
    \label{eq: Modular commutator J(U, V, W) - 3}
\end{eqnarray}
where in the first line, we used a decomposition of $K_{UV}$. For the first term of the second equality, we first reduce the region $(VW)_{\mathfrak{r}}$ to $(VW)_{\mathfrak{d}}$ and then extend the $(UV)_{\mathfrak{bd}_1}$ to $(UV)_{\mathfrak{B}}$.

Now that we have computed both lines of Eqn.~\eqref{eq: Modular commutator J(U, V, W) - 1}, let us collect those terms together. By combining Eqn.~\eqref{eq: Modular commutator J(U, V, W) - 2} and Eqn.~\eqref{eq: Modular commutator J(U, V, W) - 3}, we obtain the following decomposition for the $i$'th tri-junction:
\begin{align}
    J(U, V, W) = & 
    J(U_{\mathfrak{B}_{i}}, W_{\mathfrak{B}_{i}}, W_{\mathfrak{B}_{i}})
    + J(U_{\mathfrak{r}_{i}}, V_{\mathfrak{r}_{i}}, W_{\mathfrak{r}_{i}}) 
    \nonumber \\ 
    & - J(U_{\mathfrak{d}_{i}}, V_{\mathfrak{d}_{i}}, W_{\mathfrak{d}_{i}}).
    \label{eq: Modular commutator J(U, V, W) - 4}
\end{align}
The only difference between Eqn.~\eqref{eq: Modular commutator J(U, V, W) - 4} and the additivity formula is [Eqn.~\eqref{eq: Sum over tri-junctions}] is the presence of an additional term $J(U_{\mathfrak{d}_{i}}, V_{\mathfrak{d}_{i}}, W_{\mathfrak{d}_{i}})$.

We note that this additional term is zero:
\begin{equation}
    J(U_{\mathfrak{d}_{i}}, V_{\mathfrak{d}_{i}}, W_{\mathfrak{d}_{i}})= 0.
    \label{eq: the modular commutator of an annulus}
\end{equation}
This is because the two modular Hamiltonians over $(UV)_{\mathfrak{d}_{i}}$ and $(VW)_{\mathfrak{d}_{i}}$ can be written as a linear combination of local terms, whose resulting modular commutators are zero. (The modular commutators are zero due to the quantum Markov chain structure.) Thus, we conclude Eqn.~\eqref{eq: the modular commutator of an annulus}. In particular, we obtain
\begin{equation}
    J(U,V,W) \! = \! J(U_{\mathfrak{B}_{i}} \!, W_{\mathfrak{B}_{i}} \!, W_{\mathfrak{B}_{i}})
    \! + \! J(U_{\mathfrak{r}_{i}} \!, V_{\mathfrak{r}_{i}} \!, W_{\mathfrak{r}_{i}}).
\end{equation}
This procedure can be repeatedly applied for every tri-junction of $J(U, V, W)$, yielding the additivity formula:
\begin{equation*}
    J(U, V, W) \!=\!  J(U_{\mathfrak{r}}\!, V_{\mathfrak{r}}\!, W_{\mathfrak{r}}) \! + \! \sum_{i} J(U_{\mathfrak{B}_{i}}\!, V_{\mathfrak{B}_{i}}\!, W_{\mathfrak{B}_{i}}),
\end{equation*}
where $\mathfrak{r} = \cup_{i} \mathfrak{r}_{i}$.

Below, we provide some remarks.
First, Eqn.~\eqref{eq: Sum over tri-junctions} of the modular commutator holds even for incomplete tri-junctions because the observation used in the proof remains valid for incomplete junctions.
Second, $J(U, V, W)$ is invariant under smooth bulk deformations away from the tri-junction. 
Here, smooth bulk deformation means deformations of subsystems within the modular commutator that do not add or remove the tri-junction.
Finally, the modular commutator for the bulk residual term vanishes
\begin{equation*}
    J(U_{\mathfrak{r}}, V_{\mathfrak{r}}, W_{\mathfrak{r}}) = 0,
\end{equation*}
when the subsystems $(UV)_{\mathfrak{r}}$ and $(VW)_{\mathfrak{r}}$ are disk-shaped regions. 
This follows from the observation in Ref.~\cite{kim2022modular} that the modular Hamiltonian for a disk-shaped bulk region can be expressed as a sum of local modular Hamiltonians. 
Therefore, for the disk-like regions, the modular commutator $J(U_{\mathfrak{r}}, V_{\mathfrak{r}}, W_{\mathfrak{r}})$ can be expressed as a sum of local modular commutators. 
Each of these commutators is a quantum Markov chain and always vanishes.

\section{Geometric additivity of modular commutator for the incomplete disk}\label{sec: Incomplete disk}

In this Section, we discuss another intriguing application of the additivity formula [Eqn.~\eqref{eq: Sum over tri-junctions}], the \textit{incomplete disk}~\cite{fan2022entanglement}. 
Let us consider a disk $ABCD$ shown in Fig.~\ref{fig: SM_3}(a).
Unlike previous cases, subregion $A, B, C$ does not fully surround its tri-junction.
We also call such tripartitions $ABC$ as the incomplete disk.

\subsection{Summary of the results}

\begin{figure}[]
    \centering
    \includegraphics[width=0.9\columnwidth]{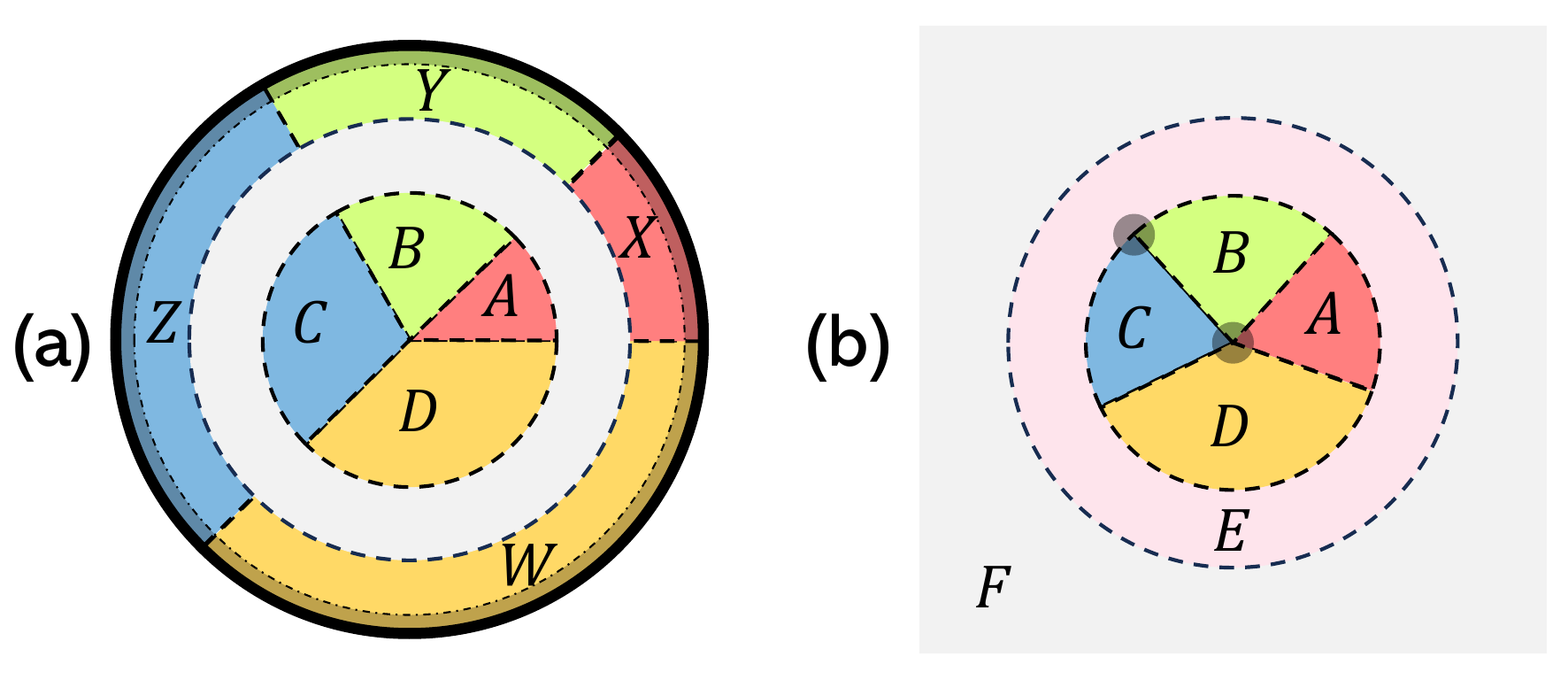}
    \caption{Schematic descriptions for incomplete tri-junction. 
    (a) Bulk disk $ABCD$ and edge annulus $XYZW$.
    (b) Partitions $(A, B, C, D, E, F)$ for the proof of Eqn.~\eqref{eq: the sum of two incomplete disks}.}
    \label{fig: SM_3}
\end{figure}

For incomplete disks $ABC$ and $BCD$, we find another type of additivity for modular commutators:
\begin{equation}
    J(A, B, C) + J(B, C, D) = \frac{\pi}{3} c_{-}.
    \label{eq: the sum of two incomplete disks}
\end{equation}
As an immediate consequence, four modular commutators of incomplete disk $ABCD$ have the following relations: $J(A, B, C) = J(C, D, A)$ and $J(B, C, D) = J(D, A, B)$.

To prove it, let us consider partitions $BCDE$ illustrated in Fig.~\ref{fig: SM_3}(b).
We first switch the modular Hamiltonian of the subsystem $CDE$ to $ABF$ using the complementary property, $K_{CDE}| \Psi \rangle = K_{ABF} | \Psi \rangle$,
\begin{align}
    J(DE, C, B) = J(AF, B, C) = J(A,B,C).
\end{align}
where in the second equality, we use $K_{ABF} = K_{AB} +  K_{F}$.
Then, we apply the additivity formula to $J(DE, C, B)$, resulting in 
\begin{eqnarray}
     J(DE,C,B) 
     = &  J(E_{\mathfrak{B}}, C_{\mathfrak{B}}, B_{\mathfrak{B}}) + J & (D_{\mathfrak{B}}, C_{\mathfrak{B}}, B_{\mathfrak{B}})  \nonumber \\
    & +  J((DE)_{\mathfrak{r}}, C_{\mathfrak{r}}, B_{\mathfrak{r}}), &
    \label{eq: decomposition of J(B, C, D)}
\end{eqnarray}
where two terms in the first line are nonvanishing terms, which are evaluated as follows
\begin{subequations}
    \begin{align}
        J(E_{\mathfrak{B}}, C_{\mathfrak{B}}, B_{\mathfrak{B}}) = & \frac{\pi}{3} c_{-}, \\
        J(D_{\mathfrak{B}}, C_{\mathfrak{B}}, B_{\mathfrak{B}}) = & J(D, C, B).
    \end{align}
\end{subequations}
The residual term in Eqn.~\eqref{eq: decomposition of J(B, C, D)} has no triple point, and the modular Hamiltonian $K_{(BC)_{\mathfrak{r}}}$ is local, but $K_{(CDE)_{\mathfrak{r}}}$ is non-local. However, it can be easily verified
\begin{equation}
    J((DE)_{\mathfrak{r}}, C_{\mathfrak{r}}, B_{\mathfrak{r}}) = 0,
\end{equation}
once we appropriately decompose $K_{(BC)_{\mathfrak{r}}}$ into smaller three chunks by Eqn.~\eqref{eq: the Petz's relation}.

In conclusion, we obtain the geometric additivity of a modular commutator for the incomplete disk,
\begin{equation}
    J(A, B, C) + J(B, C, D) = \frac{\pi}{3} c_{-}.
    \label{eq: relation of two modular commutators on the incomplete disk}
\end{equation}
It states that the sum of two incomplete modular commutators is $\frac{\pi}{3}c_-$ if the union of tripartitions forms a complete disk.

\begin{figure}[]
    \begin{center}
        \includegraphics[width=\columnwidth]{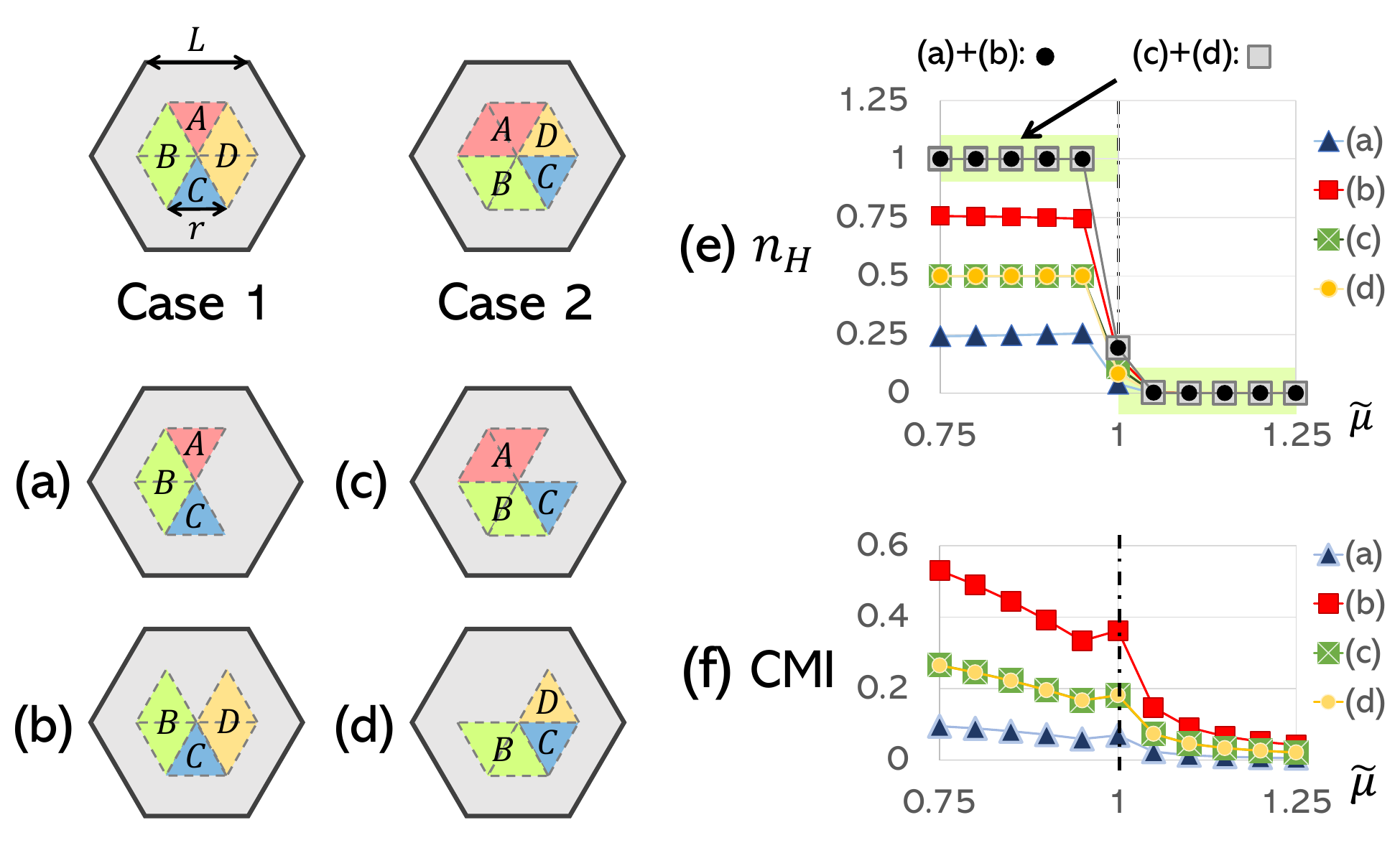}
        \caption{Numerical evaluation of modular commutator for the incomplete disks. 
        (a)-(d) The lattice realization of the incomplete disks.
        We fix the linear size of the system $L$, and the subsystem size $r$: $L=22$ and $r=15$.
        (e) Plot of $n_{H} \equiv \frac{3}{\pi} J$ of the Haldane model and the sum for the incomplete disks with tuning parameter $\tilde{\mu}$.
        (f) Plot of conditional mutual information (CMI) $I(A:C|B)$ for the incomplete disk with tuning parameter $\Tilde{\mu}$. 
        }
    \label{fig: SM_4}
    \end{center}
\end{figure}

\subsection{Numerical Calculations and Remarks}

We now test the formula \eqref{eq: the sum of two incomplete disks} for the Haldane model discussed in the main text.
We choose two cases of disks $ABCD$, which are illustrated in Fig.~\ref{fig: SM_4}(a-b) and (c-d).
For each case, the modular commutators are computed in the parameter range of $0.5 < \tilde{\mu} < 1.25$, as shown in Fig.~\ref{fig: SM_4}(e). In both cases, the sum of two modular commutators of the incomplete disks yields $\frac{\pi}{3} c_{-}$, which is consistent with our arguments.

Below, we provide some remarks. 
The first remark is about the second case in Fig.~\ref{fig: SM_4}. In the Chern insulator phase, (a) and (b) have different values. However, (c) and (d) are nearly identical despite the fact that two incomplete disks ($ABC$ and $BCD$) are different. 
It indicates that we can extract an exact chiral central charge from an incomplete disk alone, yielding $\pi c_{-} / 6$. This is due to the symmetry of the Haldane model, which enforces the following relation:
\begin{equation*}
    J(A,  B, C) = J(B,  C, D)  =  J(C, D, A)  =  J(D, A, B). 
    \label{eq: half CCC - 2}
\end{equation*}
Specifically, if the partition of incomplete disks $ABCD$ is given as below,
\begin{center}
    \includegraphics[width=0.9\columnwidth]{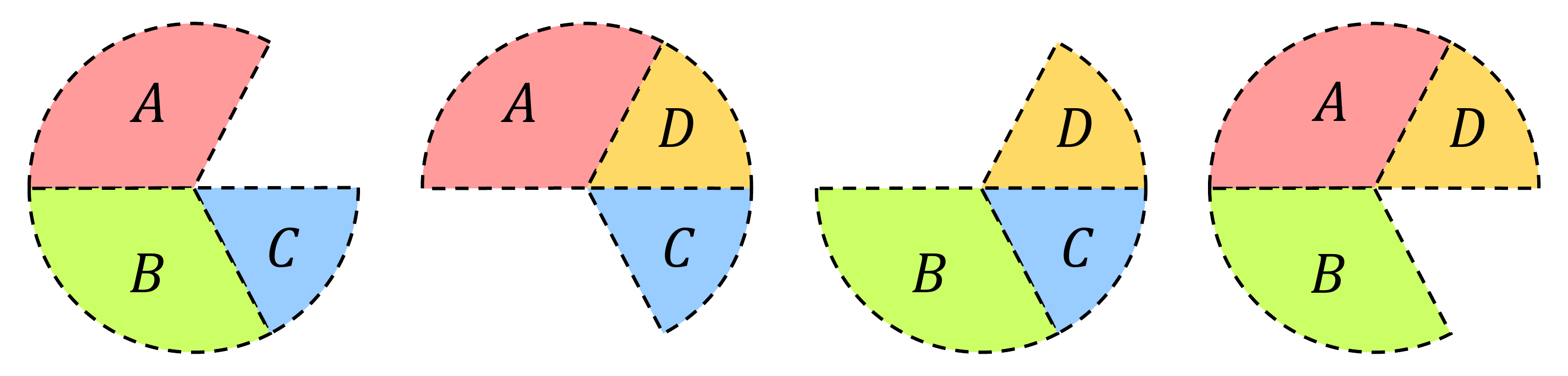}
\end{center}
and the system has a proper spatial symmetry, the two modular commutators $J(A, B, C)$ and $J(D, A, B)$ may be equal, resulting in $J(A, B, C)= J(B, C, D) = \frac{1}{2} \times \frac{\pi}{3} c_{-}$. This observation may be useful for extracting the chiral central charge more efficiently in numerical studies.

The second remark is about the incomplete tri-junction and the conditional mutual information. 
It is well-known that if a tripartite state $\rho_{ABC}$ is a quantum Markov chain, then the modular commutator vanishes, 
\begin{equation}
    I(A:C|B)_{\rho} = 0 \Longrightarrow J(A, B, C)_{\rho} = 0.
\end{equation}
Meanwhile, the tri-junction of an incomplete disk $ABC$, which makes a non-smooth boundary, provides a non-vanishing value of the conditional mutual information.
Without the tri-junction, the conditional mutual information (CMI) of the incomplete disk becomes zero because of the area law, and the modular commutator is zero.
We present the conditional mutual information of incomplete disks in Fig.~\ref{fig: SM_4}(f), which are nonzero in the chiral topological phase and decrease in the trivial phase.

Lastly, one can view Eqn.~\eqref{eq: the sum of two incomplete disks} as a bulk analog of an identity that is satisfied by the edge modular commutator~\cite{zou2022modular, kim2024conformal, fan2022entanglement}. For instance, the modular commutators at the contiguous edge intervals $XYZ$ in Fig.~\ref{fig: SM_3}(a) also satisfy the similar relation under exchanging $XYZ$ to $YZW$: $J(X, Y, Z) = \frac{\pi}{3} c_{-} - J(Y, Z, W)$. A similar observation was made in Ref.~\cite{kim2024conformal}.

\section{Additional numerical calculations for \texorpdfstring{$\pi$}{TEXT}-flux model}\label{Sec: Numerical Simulation on pi-flux}

\begin{figure}[!b]
    \centering
    \includegraphics[width=\columnwidth]{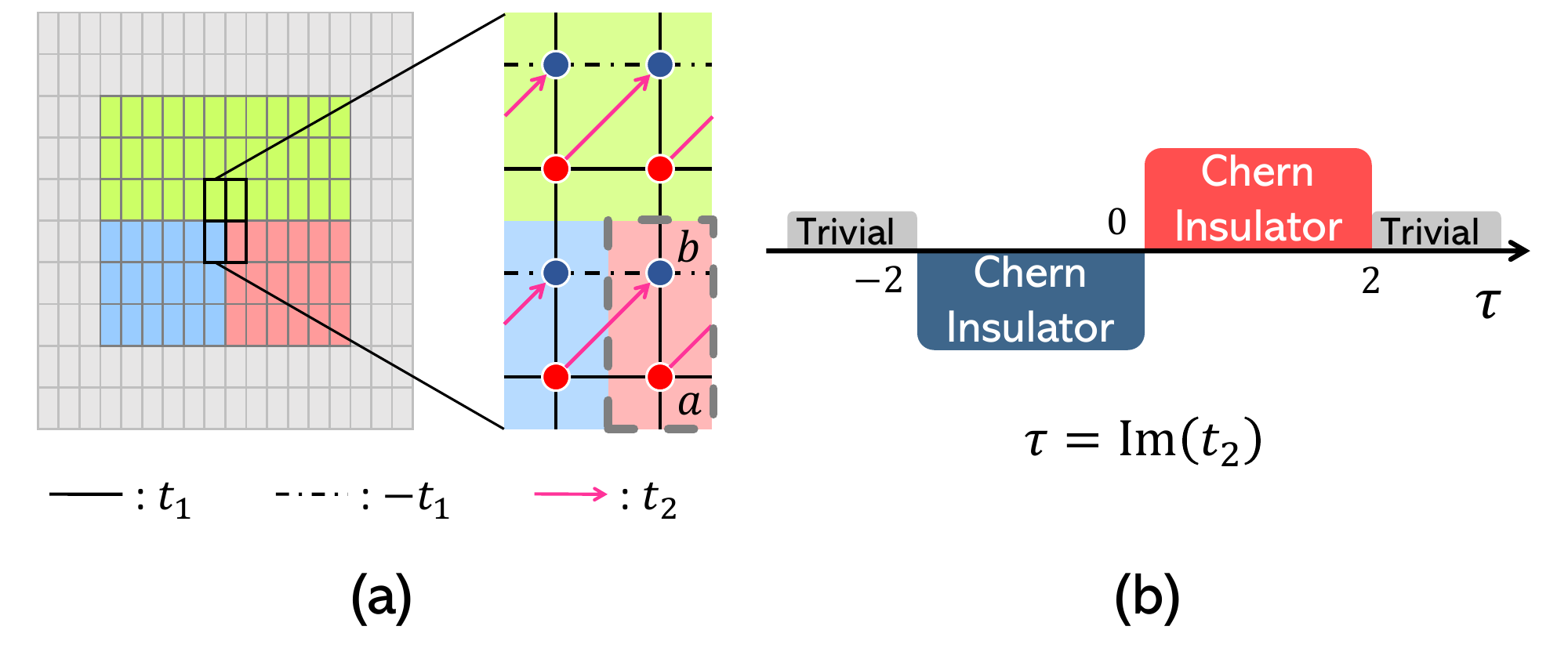}
    \caption{Schematic description of $\pi$-flux model.
    (a) Each rectangle denotes the unit cell. 
    See the main text for the details of hopping parameters.
    (b) Phase diagram of the $\pi$-flux model along $\tau = \textrm{Im}(t_{2})$-line.
    In topological phases, the chiral central charge is $c_{-}=1$ at $0<\tau<2$, and $c_{-}=-1$ at $-2<\tau<0$.
    } 
    \label{fig: SM_5}
\end{figure}

This Section provides additional numerical calculations involving the $\pi$-flux model, which is a free fermion model. Since the ground state of the free fermion systems is the Gaussian state, the reduced density matrix and modular commutators are completely determined by their two-point correlation functions~\cite{peschel2009reduced}.

\subsection{Model Hamiltonian}

\begin{figure}[]
    \centering
    \includegraphics[width=\columnwidth]{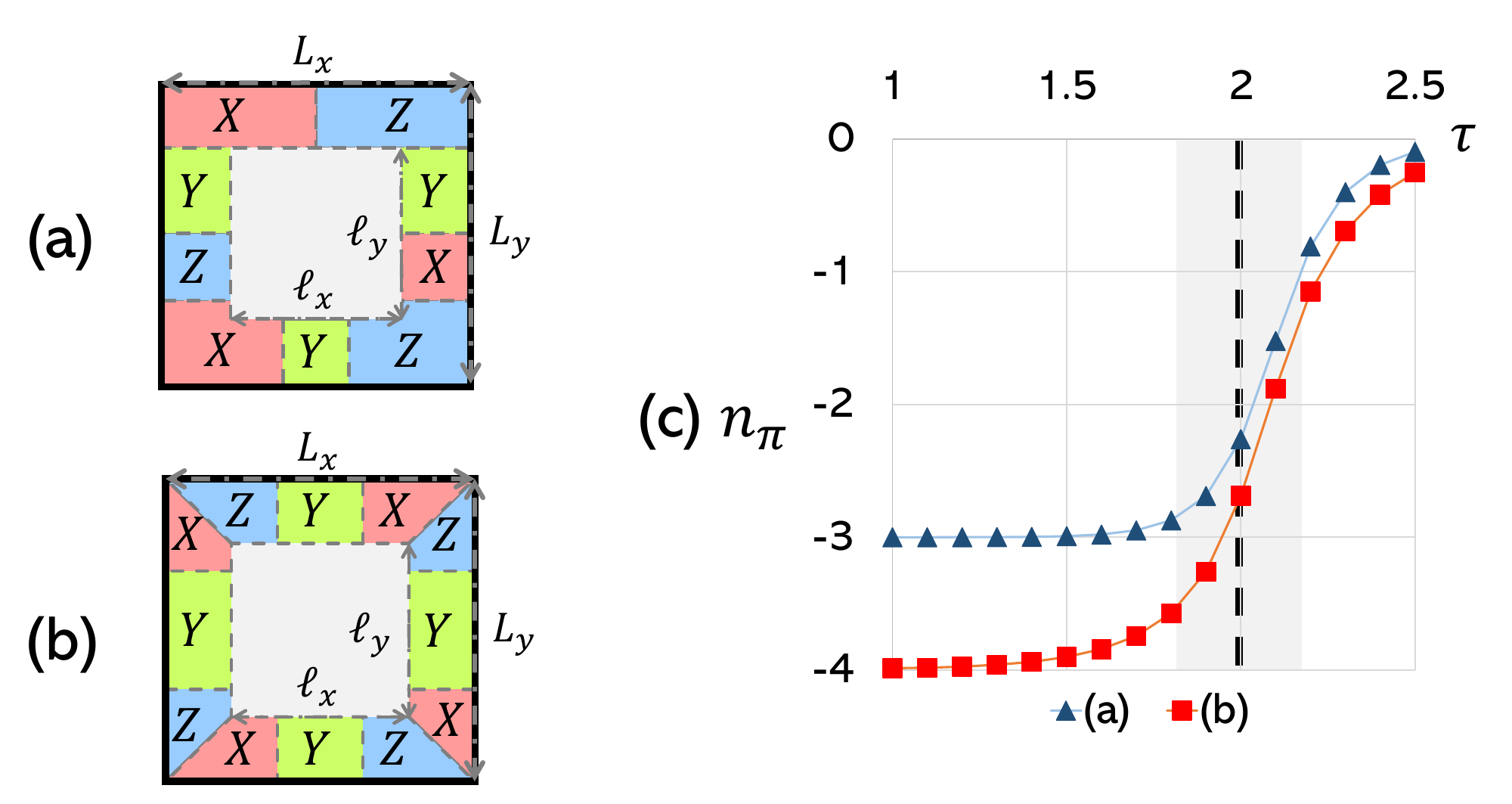}
    \caption{Numerical evaluation of the geometric integer ($n\ge 3$) for the edge pizza partition.
    (a)-(b) The lattice realization of the edge pizza portions with $n=3,4$, respectively.
    We fix the system size $L_{x} = 47$, $L_{y} = 46$, and the $\ell_{x} = 27$, $\ell_{y} = 26$.
    (c) Plot of $n_{\pi}$ for the edge pizza partition with the tuning parameter $\tau$.}
    \label{fig: SM_6}
\end{figure}

The $\pi$-flux models are defined on the square lattice, consisting of two sublattices, $a$ and $b$. 
We use $j = (r, s)$ to label the unit cell $r$ and the two sublattices $s=a,b$. 
The $\pi$-flux model is described by the following Hamiltonian:
\begin{equation}
    H_{\pi} = H_{1}(t_{1}) + H_{2}(t_{2}; \phi)
\end{equation}
The first part concerns the nearest hopping Hamiltonian with amplitude $t_{1}$, $H_{1}(t_{1}) = t_{1} \sum_{\langle j, k \rangle} \eta_{jk} c^{\dagger}_{j} c_{k}$, which is illustrated as black links.  
We use black dashed links for hopping between two $b$ sublattices, $\eta_{jk} = -1$, and black solid links for all the rest of hopping, $\eta_{jk} = 1$.
The second part includes the next nearest hopping terms with complex amplitude $t_{2}$, $H(t_{2}; \phi) = \tau \sum_{r} (i c^{\dagger}_{r+\hat{x}, b} c_{r, a} + \text{h.c} ) $, which are depicted as arrows in Fig.~\ref{fig: SM_5}(a). 
Here, we choose the amplitude to be purely imaginary $t_{2} = i \tau$.

In the $\pi$-flux model, we keep $t_{1}$ fixed at 1 and vary $\tau$. 
The model exhibits two chiral topological and trivial phases along the $\tau$-parametric path. 
The topological phase at $0 < \tau < 2$ has a chiral central charge of $c_{-} = 1$, while the other topological phase at $-2 < \tau < 0$ has opposite chiralities with a chiral central charge of $c_{-} = -1$. 
Outside of these ranges, the $\pi$-flux model is trivial. 

\subsection{Numerical results}

We first consider the pizza partitions in the $\pi$-flux model with the geometric integer $(n \ge 3)$. 
Since the model is defined on a lattice, verifying higher values of $n$ in the bulk subsystem is not straightforward. 
For this reason, we only provide the numerical evaluations of the geometric integer for the edge pizza partitions, $n_{\pi} \equiv - \frac{3}{\pi} J(X, Y, Z)$ ($n \ge 3$). 
We select $XYZ$ as shown in Fig.~\ref{fig: SM_6}(a)-(b). 
Particularly for $n=3$, we choose the lengths of each interval, $X$, $Y$, and $Z$, not equal.
Numerical estimations of the geometric integers $n_{\pi}$ are presented in Fig.~\ref{fig: SM_6}(c). 
The geometric integers well-converge for a large energy bulk gap ($1<\tau<1.5$).
As $\tau$ approaches the critical point, the values of the modular commutator start to break down due to the decreasing energy gap.
This is because to be well-converged to the integer, each interval demands a larger size than the bulk correlation length.

\begin{figure}[t]
    \centering
    \includegraphics[width=\columnwidth]{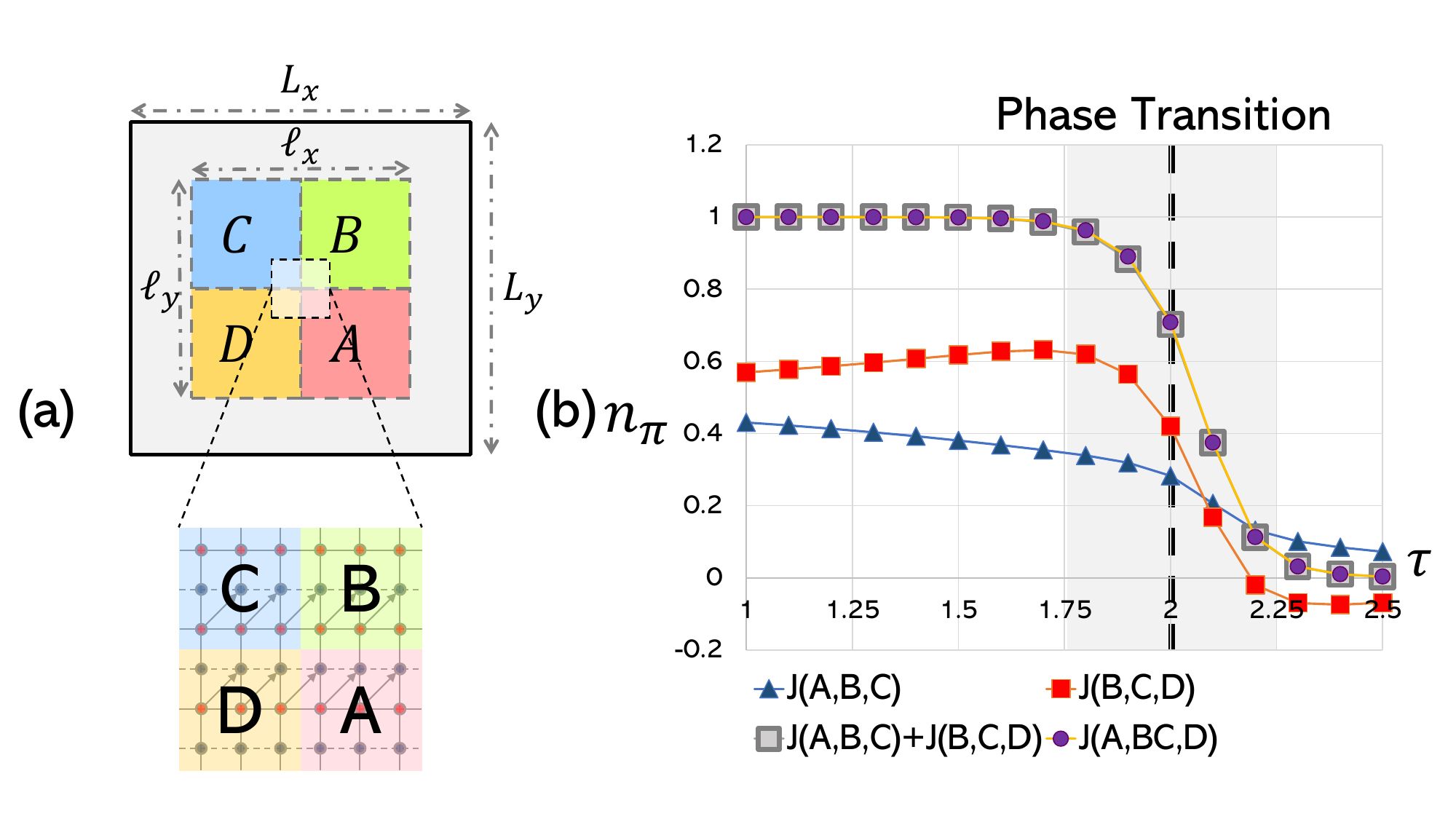}
    \caption{Numerical evaluation of the modular commutator for the incomplete disk $ABCD$.
    (a) Incomplete disk $ABCD$ in $\pi$-flux model. 
    We fix the disk size $(\ell_{x}, \ell_{y}) = (15, 14)$ and the system size $(L_{x}, L_{y}) = (47, 46)$.
    (c) Plot of modular commutators and the sum for the incomplete disks with the tuning parameter $\tau$.}
    \label{fig: SM_7}
\end{figure}

Next, we consider the $\pi$-flux model shown in Fig.~\ref{fig: SM_7}.
We choose a partition $ABCD$ of the $\pi$-flux model as depicted in Fig.~\ref{fig: SM_7}(a). 
We compute the modular commutators at $1 < \tau < 2.5$ as shown in Fig.~\ref{fig: SM_7}(b). 
Two modular commutators of the incomplete disks exhibit variations as the parameter $\tau$ changes.
However, the sum always equals $J(A, BC, D),$ in both chiral topological and trivial phases. 
This is consistent with our discussion.

\section{Absence of the residual term in the invertible states and remarks on the non-invertible states}\label{sec: absence of the residual term}

In this Section, we prove the residual term of the additivity formula is absent for the invertible bulk. 
We also discuss the additivity for the non-invertible system.

\subsection{Proof: Absence of the residual term in the invertible bulk}

The main issue in evaluating the modular commutator of the pizza partition is that it has a non-trivial topology. 
For example, the subsystem of the bulk pizza partition has a bowtie shape ($\bowtie$); it is non-trivial whether it is a connected one region or disconnected two regions. 
These issues become subtle when the system is topologically ordered. 
However, based on the observations below, this issue can be circumvented for the invertible bulk.

\begin{figure}[]
    \centering
    \includegraphics[width=\columnwidth]{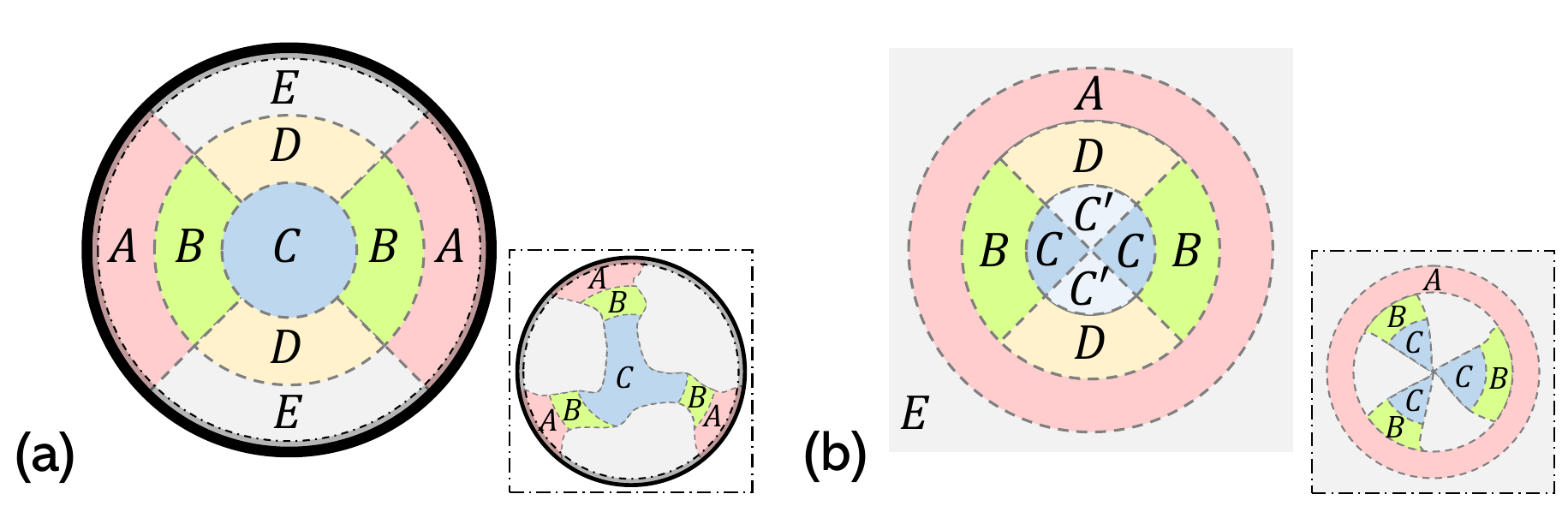}
    \caption{Multipartite regions in a two-dimensional system. 
    If the bulk is invertible, the conditional mutual information $I(A: C|B)$ vanishes for any smooth bulk deformation of $ABC$.
    (a) $A, E$ are disconnected edge region at physical boundary, and $BCD$ is bulk region. 
    (b) A multipartite region $ABCC'D$ in the bulk.}
    \label{fig: SM_8}
\end{figure}

For concreteness, let us consider a two-dimensional gapped system $ABCDE$ on a disk as shown in Fig.~\ref{fig: SM_8}(a). 
The entanglement entropies of these subsystems are related as
\begin{equation}
    I(A:C|B) + I(E:C|D) = \Delta(B, C, D),
    \label{eq: long subsystem}
\end{equation}
where $\Delta(B, C, D) = S_{BC} + S_{CD} - S_{B} - S_{D}$ and we use $S_{AB} = S_{CDE}$ and $S_{ABC} = S_{DE}$ for pure state.
The bulk subsystem $BCD$ is similar to the partition for the axiom $\textbf{A1}$ [Eqn.~\eqref{eq: Axiom A1}], but the subsystem $B$ and $D$ are the union of the disconnected two regions.
In this case, $\Delta(B, C, D)$ is no longer zero; rather, it gives the topological entanglement entropy ($\gamma_{\text{top}}$)~\cite{kitaev2006topological, levin2006detecting}: $\Delta (B, C, D) = 2 \gamma_{\text{top}}$. 
Note that the topological entanglement entropy contribution in $\Delta$ depends on the number of the disconnected components of $B$ (or $D$). 
For instance, if the $B$ and $D$ are the union of $m$ disconnected region, then $\Delta(B, C, D) = 2(m-1) \gamma_{\text{top}}$.

A key observation of Eqn.~\eqref{eq: long subsystem} is that if the bulk is invertible ($\gamma_{\text{top}} = 0$), the quantum state on the subsystem $ABC$ is the quantum Markov state. 
Thus the modular Hamiltonian of $ABC$ is local~\cite{petz2003monotonicity}:
\begin{equation}
    K_{ABC} = K_{AB} + K_{BC} - K_{B}.
    \label{eq: local modular Hamiltonian of the pizza partition}
\end{equation}
This observation can be further generalized. 
For example, if the bulk is invertible, we have $I(A: C|B) = 0$ for the partition $ABC$ shown in the inset of Fig.~\ref{fig: SM_8}(a), where the edge region $A$ and the bulk region $B$ consist of three disconnected regions.

One can also obtain similar results for the bulk multipartition $ABCC'D$ in Fig.~\ref{fig: SM_8}(b). 
We find 
\begin{equation}
    I(A:C|B) + I(A:C'|D) = \Delta(B,CC',D).
    \label{eq: quasi n-hole disk}
\end{equation}
Here, $\Delta(B, CC', D) = 2\gamma_{\text{top}}$ and we use the fact that the state in sufficiently distant regions is a product state, for example, $S_{AD} = S_{BCC'} + S_{E}$. 
It is clear that when the bulk is invertible ($\gamma_{\text{top}} = 0$), the modular Hamiltonian becomes local.

\begin{figure}[]
    \centering
    \includegraphics[width=\columnwidth]{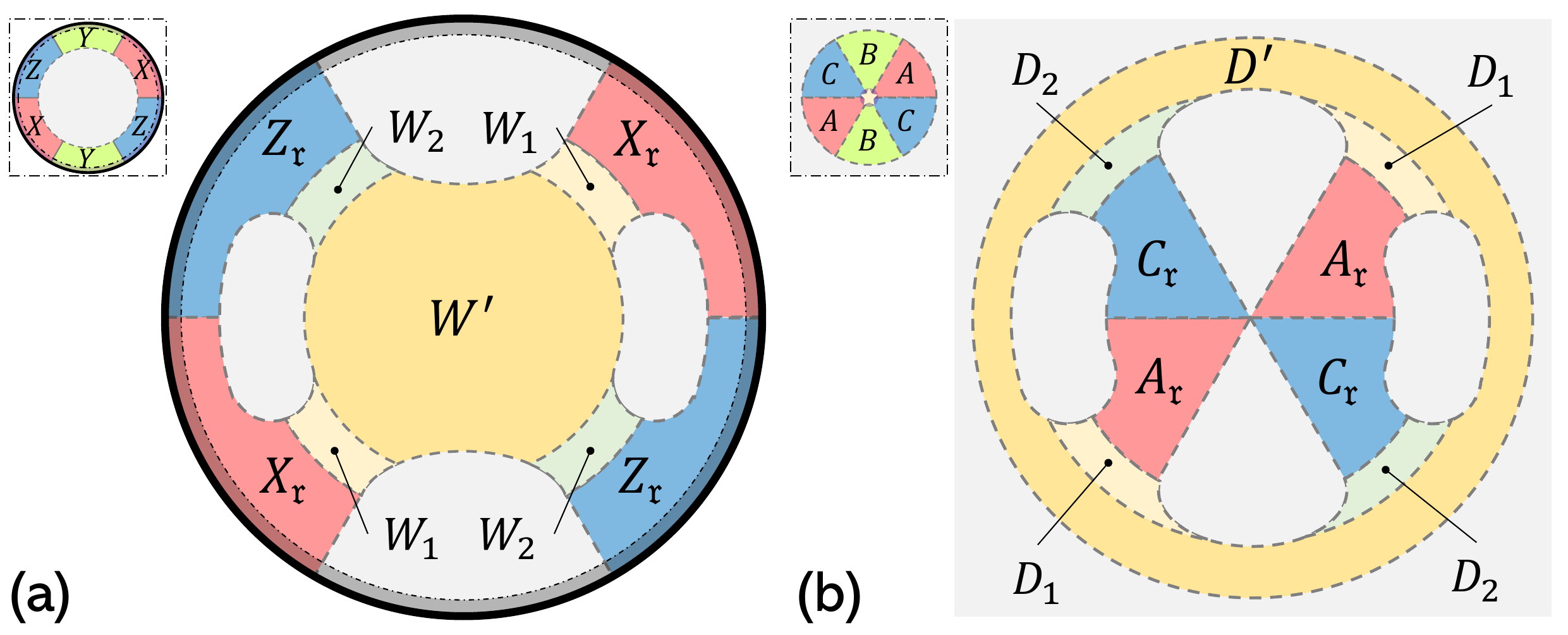}
    \caption{Partitions of the residual term, $J(Z_{\mathfrak{r}}, W_{\mathfrak{r}}, X_{\mathfrak{r}})$ and  $J(C_{\mathfrak{r}}, D_{\mathfrak{r}}, A_{\mathfrak{r}})$.}
    \label{fig: SM_9}
\end{figure}

Below, we prove the absence of the residual term. 
While we focus on the examples in Fig.~\ref{fig: SM_9}, the proof for the other cases is straightforward as well.

Let us first consider the edge residual term $J(Z_{\mathfrak{r}}, W_{\mathfrak{r}}, X_{\mathfrak{r}})$ and let $W_{\mathfrak{r}} = W_{1}W_{2}W'$ as shown in Fig.~\ref{fig: SM_9}(a). 
From Eqn.~\eqref{eq: local modular Hamiltonian of the pizza partition}, we have 
\begin{align*}
    K_{(ZW)_{\mathfrak{r}}} = K_{Z_{\mathfrak{r}}W_{2}} + K_{W_{\mathfrak{r}}} - K_{W_{2}} \\
    K_{(XW)_{\mathfrak{r}}} = K_{X_{\mathfrak{r}}W_{1}} + K_{W_{\mathfrak{r}}} - K_{W_{1}}
\end{align*}
and, by plugging into $J(Z_{\mathfrak{r}}, W_{\mathfrak{r}}, X_{\mathfrak{r}})$, we obtain
\begin{align*}
    J(Z_{\mathfrak{r}}, W_{\mathfrak{r}}, X_{\mathfrak{r}}) 
    = & i \langle [K_{(ZW)_{\mathfrak{r}}}, K_{(WX)_{\mathfrak{r}}}] \rangle \\
    = & J(Z_{\mathfrak{r}}, W_{2}, W_{\mathfrak{r}} \setminus W_{2}) + J(W_{\mathfrak{r}} \setminus W_{1} , W_{1}, X_{\mathfrak{r}}) \\
    = & 0.
\end{align*}
In the second and last line, we use the two properties of the modular commutator in Ref.~\cite{kim2022modular}. 
In the second line, we use that if any of the subsets of the modular commutator is empty, the modular commutator vanishes:
\begin{equation}
    J(\varnothing, B, C) = J(A, \varnothing, C) = J(A, B, \varnothing) = 0.
\end{equation}
In the last line, we use that if the underlying state is a quantum Markov state, the modular commutator vanishes: 
\begin{equation}
    I(A:C|B) = 0 \quad \Longrightarrow \quad J(A, B, C) = 0.
\end{equation}

We can also prove the absence of the bulk residual term similarly.
Consider the bulk residual term $J(C_{\mathfrak{r}}, D_{\mathfrak{r}}, A_{\mathfrak{r}})$ in Fig.~\ref{fig: SM_9}(b) and let $D_{\mathfrak{r}} = D_{1}D_{2}D'$.
Then, we have
\begin{align*}
    J(C_{\mathfrak{r}}, D_{\mathfrak{r}}, A_{\mathfrak{r}}) 
    = & i \langle [K_{(CD)_{\mathfrak{r}}}, K_{(DA)_{\mathfrak{r}}}] \rangle \\
    = & J(C_{\mathfrak{r}}, D_{2}, D_{\mathfrak{r}} \setminus D_{2}) + J(D_{\mathfrak{r}} \setminus D_{1} , D_{1}, A_{\mathfrak{r}}) \\
    = & 0,
\end{align*}
where the properties of the modular commutator and the decomposition of the modular Hamiltonian are used:
\begin{align*}
    K_{(CD)_{\mathfrak{r}}} =& K_{C_{\mathfrak{r}}D_{2}} + K_{D_{\mathfrak{r}}} - K_{D_{2}} \\
    K_{(AD)_{\mathfrak{r}}} =& K_{A_{\mathfrak{r}}D_{1}} + K_{D_{\mathfrak{r}}} - K_{D_{1}}.
\end{align*}
We emphasize that this argument for the invertible bulk is applicable quite generally. 
Therefore, the additivity of the modular commutator manifests as long as the bulk is invertible and the subsystem is sufficiently larger than the bulk correlation length.

\subsection{Discussion: non-invertible bulk}

In a topologically ordered system, there is in fact a state for which the additivity of the modular commutator does hold. 
In the entanglement bootstrap program~\cite{shi2020fusion}, certain \emph{merged states} can be such states. 
A merged state is a maximum entropy state locally consistent with the ground state (or vacuum state, $\sigma$).
In anyon theory, a merged state can be considered as a maximum entropy mixture containing anyons.

\begin{figure}[!t]
    \centering
    \includegraphics[width=\columnwidth]{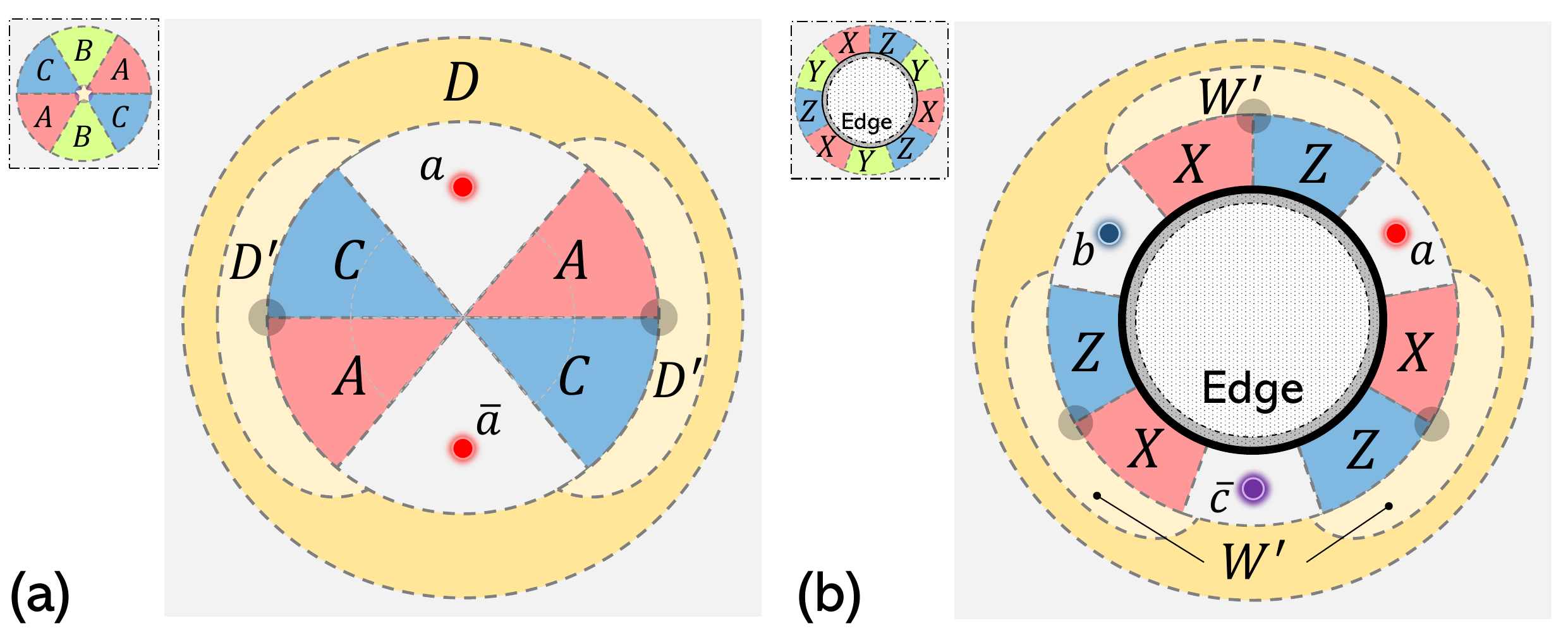}
    \caption{The modular commutator of the pizza partitions in the non-invertible system.
    (a) An ensemble of the states containing anyons ($a, \bar{a}$), where $\bar{a}$ is anti-charge of topological charge $a$.
    Anyons are located in the subsystem $B$.
    (b) An maximum entropy ensemble of anyons ($a, b, \bar{c}$) that fuse into vacuum, where $a$ and $b$ fuse into $c$. 
    The state $\tilde{\tau}$ satisfies $I(ZX: W \setminus W' | W')_{\tilde{\tau}} = 0$.}
    \label{fig: SM_10}
\end{figure}

For example, consider a merged state on the bulk subsystem $CDA$ as shown in Fig.~\ref{fig: SM_10}(a).  
A merged state of our interest, say $\tau$, is a maximum entropy state such that
\begin{equation}
    I(CA:D \setminus D' | D')_{\tau} = 0.
\end{equation} 
In entanglement bootstrap, such a state can be generated by merging the states on disk and annulus, whose resulting subsystem is the two-hole disk~\footnote{
Note that the subsystem $CDA$ is not a true two-hole disk, and its information convex set is not well-defined due to its bowtie shape ($\bowtie$). 
%To be $CDA$ a true two-hole disk, the thickness should be larger than the correlation length.
However, conditional independence remains true because of the strong subadditivity.
}.
In the context of anyon theory, an ensemble on $CDA$ that has anyons ($a, \bar{a}$) in the holes ($B$) with probability $p_{a\bar{a}}^{1} = d^{2}_{a} / \mathcal{D}^{2}$ satisfies such conditional independence.
Here, $d_{a}$ is quantum dimension of anyon $a$ and $\mathcal{D}= \sqrt{\sum_{a \in \mathcal{C}} d_{a}^{2}}$ is total quantum dimension over anyon superselection sector $\mathcal{C}$.
% To be precise, let us denote $a \in \mathcal{C} = \lbrace 1, a, b, \cdots \rbrace$ as topological charge of anyons, $d_{a}$ as quantum dimension and $\mathcal{D}= \sqrt{\sum_{a \in \mathcal{C}} d_{a}^{2}}$ as total quantum dimension.
% Then, a mixture of states ($\rho^{a\bar{a}1}$) that anyons ($a, \bar{a}$) fuse into a vacuum `1' with probability $p_{a\bar{a}}^{1} = d^{2}_{a} / \mathcal{D}^{2}$ over anyon superselction sector $\mathcal{C}$ satisfies such conditional independence [Fig.~\ref{fig: SM_10}(a)]. 

For this merged state ($\tau$), the modular commutator $J(C, D, A)_{\tau}$ is 
\begin{align}
    J(C, D, A)_{\tau}  
    = & J(C, D', A)_{\sigma} \nonumber \\ 
    = & 2 \times \frac{\pi}{3} c_{-} + J(C_{\mathfrak{r}}, D'_{\mathfrak{r}}, A_{\mathrm{r}})_{\sigma}, 
\end{align}
where we use the conditional independence of the maximum entropy state, $I(C:D \setminus D' | D')_{\tau} = 0$ and $I(A:D \setminus D' | D')_{\tau} = 0$, in the first line and the additivity formula in the second line. 
The residual term vanishes $J(C_{\mathfrak{r}}, D'_{\mathfrak{r}}, A_{\mathrm{r}})_{\sigma} = 0$ because the subsystems are disk-like region. 
Thus, we have $J(C, D, A)_{\tau} = 2 \times \frac{\pi}{3} c_{-}$.

Similarly, the additivity of the modular commutator manifests for a maximum entropy state, say $\tilde{\tau}$, on the edge pizza partition. 
For example, consider the subsystem $ZWX$ on a disk as shown in Fig.~\ref{fig: SM_10}(b). 
%~\footnote{If the underlying manifold has a hole, we need additional assumption, \emph{full-boundary axioms \textbf{A0}}, recently introduced in Ref.~\cite{kim2024chiral}} 
If the underlying state is a maximum entropy state such that $I(ZX: W \setminus W' | W')_{\tilde{\tau}} = 0$, we have
\begin{equation}
    J(Z, W, X)_{\tilde{\tau}} = J(Z, W', X)_{\sigma} = - 3 \times \frac{\pi}{3} c_{-}.
\end{equation}
Note that the maximum entropy state $\tilde{\tau}$ can also be expressed as an ensemble containing anyons [Fig.~\ref{fig: SM_10}(b)].
% For example, an ensemble on $ZWX$ such that anyons ($a, b$) fuse into with probability $p_{ab\bar{c}}^{1} = N_{ab\bar{c}}^{1} d_{a}d_{b}d_{\bar{c}}/ \mathcal{D}^{4}$ has the desired conditional independence, $I(ZX: W \setminus W' | W')_{\tilde{\tau}} = 0$.
% Here, $N_{ab\bar{c}}^{1}$ is the fusion multiplicity.

% Note that a state such that vacuum charge `1' splits into anyons ($a, b, \bar{c}$) with probability $p_{ab\bar{c}}^{1} = N_{ab\bar{c}}^{1} d_{a}d_{b}d_{\bar{c}} / \mathcal{D}^{4}$ has the desired conditional independence, $I(ZX: W \setminus W' | W')_{\tilde{\tau}} = 0$. 
% Here, $N_{ab\bar{c}}^{1}$ is fusion multiplicity.

Therefore, if one proves that the modular commutator is invariant even in the presence of anyons, then the geometric additivity of the modular commutator is also proven
\begin{subequations}
    \begin{align}
        J(C, D, A)_{\tau} \overset{?}{=} & J(C, D, A)_{\sigma} \\
        J(Z, W, X)_{\tilde{\tau}} \overset{?}{=} & J(Z, W, X)_{\sigma}
    \end{align}
\end{subequations}
by using the identity, $J(C, D, A)_{\sigma} = J(A, B, C)_{\sigma}$ and $J(Z, W, X)_{\sigma} = J(X, Y, Z)_{\sigma}$.
Thus, examining whether the modular commutator on the \emph{pizza partition} remains invariant in the presence of anyons may be one of the strategies for proving the additivity of the modular commutator, even in non-invertible systems. 
We leave it as future work.

\end{document}